\documentclass [12pt,aps,a4paper,prb]{revtex4}
\input{epsf}
\usepackage{graphicx}
\usepackage{amssymb}
\usepackage{latexsym}

\begin{document}
\renewcommand{\topfraction}{0.9}
\renewcommand{\textfraction}{0.1}
\renewcommand{\floatpagefraction}{0.75}
\topmargin 0.0cm
\newcommand {\be} {\begin{equation}}
\newcommand {\ee} {\end{equation}}
\newcommand {\bea} {\begin{eqnarray}}
\newcommand {\eea} {\end{eqnarray}}
\newcommand {\D} {\displaystyle}
\newcommand {\fle} { \stackrel{\rightarrow} }
\newcommand {\som} { \stackrel{\wedge} }
\newcommand {\iz} {\left}
\newcommand {\de} {\right}
\newcommand {\etal} { \emph{et al }}
\renewcommand{\theequation} {\arabic{section}.\arabic{equation}}

\setcounter{section}{0}
\begin{center}
{\Large \bf Theory of competitive counterion adsorption on flexible 
polyelectrolytes : Divalent salts}

\bigskip

{\bf Arindam Kundagrami and M. Muthukumar$^{a)}$}

\bigskip
{\it Department of Polymer Science and Engineering \\
University of Massachusetts at Amherst, Amherst, MA 01003}

\begin{abstract}
Counterion distribution around an isolated flexible polyelectrolyte in 
the presence of a divalent
salt is evaluated using the adsorption model [M. Muthukumar, J. Chem. Phys. 
{\bf 120}, 9343 (2004)] that considers Bjerrum length, salt concentration, and
local dielectric heterogeneity as physical variables in the system. Self 
consistent calculations of effective charge and size of polymer show that
divalent counterions replace condensed monovalent counterions in competitive
adsorption. The theory further predicts that at modest physical conditions,
polymer charge is compensated and reversed with increasing divalent salt. 
Consequently, the polyelectrolyte collapses and reswells, respectively. 
Lower temperatures and higher degrees of
dielectric heterogeneity enhance condensation of all species of ions.  
Complete diagram of states for the effective charge calculated as functions 
of Coulomb strength and salt concentration suggest that
(a) overcharging requires a minimum Coulomb strenth, and (b) progressively higher 
presence of salt 
recharges the polymer due to either electrostatic screening (low Coulomb 
strength) or
negative coion condensation (high Coulomb strength). A simple theory of 
ion-bridging is also presented which predicts a first-order collapse of 
polyelectrolytes.  The theoretical predictions are in agreement with generic 
results from experiments and simulations.
    
\end{abstract}

\end{center}


\maketitle

\noindent \small a)Author to whom correspondence should be addressed. E-mail:
muthu@polysci.umass.edu

\normalsize

\section{Introduction}

\setcounter{section}{1}
\setcounter{equation}{0}

Understanding charged polymers has again become a major focus of the polymer
community in the last few years. With respect to uncharged systems, there are two 
additional length scales traditionally considered in theoretical treatments of
salty polyelectrolyte solutions. One is the Bjerrum length $l_B$, which sets
the length scale for the strength of the Coulomb interaction at a particular 
temperature $T$ in a
specific solvent of dielectric constant $\epsilon$, given by
\bea
\label{Bjerrum}
l_B = \frac{e^2}{4 \pi \epsilon_0 \epsilon k_B T},
\eea         
where $e$ is the electron charge, $\epsilon_0$ is the vacuum dielectric
constant, and $k_B$ is the Boltzmann constant. The other
one is the Debye screening length $\kappa^{-1}$, which sets the length scale 
for screening due to dissociated ions, given by
\bea
\label{Debye}
\kappa^2 = 4 \pi l_B \sum_i Z_i^2 c_i,
\eea
where the sum is over all species ($i$) of mobile ions of valency $Z_i$ and
concentration $c_i$. In addition to the above two, a third length scale can 
be conceived\cite{mut-04} relating to the dielectric heterogeneity, 
which accounts for the difference in the dielectric constant 
in the vicinity of the chains and in the bulk solvent. In this paper,
we consider these three length scales to theoretically determine the effective 
charge and conformation of a single isolated polyelectrolyte chain in the 
presence of divalent salts. 

The typical non-monotonic dependence of the average conformation of flexible 
polyelectrolyte chains on temperature is now well-known. 
At very high temperatures, the chains in dilute
salt-free solutions are in their athermal states with self-avoiding-walk (SAW) 
statistics. Although the chains are
fully charged at these conditions, the electrostatic repulsion among monomers
remains negligible compared to
thermal fluctuations. As the temperature decreases, electrostatics
becomes progressively important and the chains expand due to inter-monomer
repulsion even beyond the excluded-volume swelling. At even lower temperatures, 
counterions condense on the chains sufficiently reducing the net polymer charge, 
and consequently the chains contract again.
The presence of small-molecular monovalent salts has long been known to enhance
this condensation effect, and the resulting collapse occurs at higher 
temperatures. Monovalent salts, however, collapse polyelectrolytes at 
temperatures that are still typically way below modest temperatures ( e. g., 
room temperature for aqueous solutions).  Presence of divalent (or
multivalent) salts, however, induces a drastic qualitative change in
polyelectrolyte behaviors. A modest number of divalent counterions in water can
effectively neutralize and collapse polyelectrolytes at room temperatures.
Further, addition of a higher amount of divalent salt can even reverse the
charge on the polymer (the phenomenon known as
{\it overcharging} or {\it charge inversion} or {\it charge reversal}) at
certain physical conditions.

The condensation of counterions on flexible polyelectrolyte chains has 
traditionally been covered in the Manning model\cite{man-69} originally designed
for infinitesimally thin and infinitely long rod-like molecules. However, 
Manning's argument has been found inadequate\cite{mut-04} 
for flexible polyelectrolytes, particularly for complex systems with 
multivalent ions.  Flexibility allows significant bending of molecules
due to charge compensation at lower temperatures, allowing substantial 
changes in the conformational entropy of the polymer. 
Manning's assumption that the discrete nature of the charged groups has
a secondary effect becomes entirely invalid for multivalent ions. 
It is precisely this discreteness that is found responsible for complete
charge compensation (and resulting contraction of polyelectrolytes) and subsequent 
overcharging at modest temperatures by multivalent salt counterions. This
overcharging behavior is unexplainable within the Poisson-Boltzman formalism
which considers a continuum description of the charge density. In order to
address precipitation of chains for high counterion valence, initial 
theories\cite{gonolv-95}
considered translational free energy of polyions and salt ions alongwith
screened Coulomb interaction between charges.  With prefixed values
of the excluded volume exponent $\nu$ (i.e., prefixed radius of gyration, $R_g$),
the free energy was minimized in terms of counterion species, and the correlated
multivalent ions were shown\cite{olvetal-95} to induce attraction between 
monomers (through ion-'bridging') capable of collapsing a chain. Redissolution
of chains was also observed at higher (multivalent) salt concentrations, but 
it was explained by a reduced bridging force due to electrostatic 
screening (as opposed to overcharging). Later, unscreened Coulomb interaction
within  condensed ion-pairs was first addressed\cite{kuhetal-98} without
considering the chain entropy, and the theory predicted dependencies of 
the degree of ionization, $f$, (which is the total effective charge density of  
polyelectrolytes after accounting for the condensed ions) on temperature and 
salt concentrations to be similar to Manning's argument.  A two-state (rod-like 
and collapsed) model
for condensation predicted\cite{sololv-00} that chain collapse occurs when the 
total charge of multivalent cations equals to that of the ionizable groups of 
the polymer, implying the condensation of almost {\it all} 
added multivalent ions at modest temperatures. The two-state 
theory\cite{sololv-00,sol-02} treats the collapsed state at low temperatures 
as an amorphous ionic solid similar to simple electrolytes
(say, NaCl), and, therefore, still ignores the
chain entropy and bending-related reorganization of condensed charges at low
temperatures. 

Generic experiments\cite{conhud-03,conetal-03,hub-93,ikeetal-98,schhub-00,
deletal-94} and simulations\cite{chayet-03-2} 
have shown that added cations with higher valence are more
effective in compactifying a long DNA molecule (or a polyanion
in general), and that implies a dominant electrostatic 
mechanism for polyelectrolyte collapse. Within the purview of this concept,
the issue of overcharging induced by the addition of multivalent salts was 
investigated theoretically for rodlike 
DNA\cite{nguetal-00,nguetal-00-2,nguetal-00-3} molecules. It
was further noticed\cite{schpin-98,sch-99} that there is a typical range in salt 
concentration, in
which the short-ranged attractions between monomers are effective due to the 
proximity of the
isoelectric point (at which the average effective charge of the polymer is zero). 
The total effective DNA charge was found to reverse sign if the multivalent salt
concentration was increased above this range. This concentration window, within 
which a flexible chain remains collapsed with virtually negligible net charge on 
it, is predicted to be very small in recent simulations\cite{hsilui-06}, which 
otherwise observe definite overcharging. Although rapid collapse due to charge 
neutralization for higher salt concentrations or lower temperatures is abundantly
observed in experiments\cite{beeetal-97,praetal-01,pra-05} and in  
simulations\cite{stekre-95,stekre-93,liumut-02, winetal-98}, charge reversal 
by multivalent salt is not universally observed\cite{liuetal-03}. Very recent
experiments, however, have lent support to both charge 
inversion\cite{muretal-03,besetal-07} and counterion mediated 
attraction\cite{qiuetal-07} in biological polyelectrolytes such as DNA.   
 
An effective two-parameter theory\cite{witetal-95} considered the adsorption
process of counterions and formation of mono and di-complexes between 
negatively charged monomers and divalent positively charged salt ions. The theory 
predicted charge neutralization and subsequent charge-reversal of the chain 
backbone at moderate concentrations of divalent salts. The predictions, however, 
are limited to weak polyelectrolytes without consideration to the chain energy 
and hence chain configurations at various physical conditions.    
To address the chain entropy of flexible polyelectrolytes and its
role in counterion distribution, Muthukumar 
developed\cite{mut-04} a continuum theory of counterion condensation as an
adsorption process.  Condensation in this argument is facilitated at lower 
temperatures but, unlike in previous theories, is coupled with the configurational 
free energy of the polymer.   The adsorption theory\cite{mut-04} considers 
continuous values 
for the size ($R_g$) of a single chain treated as a continous curve, which 
provide an appropriate description of entropy  of a flexible chain. The theory
treats the salt-free and salty conditions with monovalent counterions in dilute
solutions of flexible polyelectrolytes.  The parametric analysis of the 
competition bewteen the translational 
entropy of counterions and the electrostatic energy gain of condensed ions 
reproduces all classical results including the chain-collapse due to 
short-ranged dipole interactions at low temperatures\cite{winetal-98,liumut-02}.
In addition to the length scales $l_B$ (Bjerrum length) and $\kappa^{-1}$ (Debye 
length) in the
charged system, the adsorption theory uses the concept of a dielectric mismatch
parameter, $\delta$, which captures the fact that the dielectric constant has
much lower values near the chain backbone of a polyelectrolyte or protein than 
in the bulk\cite{meheic-84,lampac-97,roublo-98} solvent. In generic 
polyelectrolyte solutions, $\delta$ is the ratio of the bulk to local dielectric 
constants, and the range in which $\epsilon$ assumes its bulk value sets a new
length scale. This mismatch in 
$\epsilon$, if substantial, will create higher potential gradients that 
can electrostatically guide counterions toward oppositely charged monomers. 
The theory showed that this may significantly increase 
counterion condensation at modest temperatures leading to a lower effective 
charge and smaller
size of the polyelectrolyte. Monovalent counterions, however, was shown not
to be able to collapse a chain completely at modest temperatures (say, the
room temperature at which $l_B \sim 7\AA$ in water). 

In this paper, we extend Muthukumar's adsorption theory\cite{mut-04} to include
divalent counterions. The basic concept still relies on the competition between
the electrostatics of condensed ions and the entropy of free ions. The present 
model is intended to analyze the competitive displacement of monovalent 
counterions by divalent counterions when a salt-free dilute solution of flexible
polyelectrolytes is mixed with a salt solution of divalent ions. As a specific 
example of this situation, we consider
a single isolated polyelectrolyte (NaPSS) in a dilute solution (water) with 
monovalent
counterions of its own but with no additional monovalent salt (NaCl) in general.
We monitor the chain conformation and arrangement of condensed ions as functions 
of temperature, degree of dielectric mismatch, and concentration of divalent 
salt (BaCl$_2$). In addition to
the condensation of monovalent and divalent counterions (Na$^+$ and Ba$^{++}$,
respectively), we also take into account the attachment of negative salt 
coions (Cl$^-$), as previous theories\cite{witetal-95,sol-02} and 
simulations\cite{liuetal-03}
indicate substantial presence of negative coions near the chain backbone.
The configurational free energy of the system depends on the extent of
adsorption of various species of ions on the chain backbone as well as
on the size of the polymer, and all contributing factors remain non-trivially
coupled. These factors are assessed self-consistently with the important 
assumption of the adsorption theory being that the chemical potentials in 
the adsorbed and free states of the ions are the same. A numerical minimization
of the free energy with respect to the fraction of condensed ions and the chain 
size determines the equilibrium values of the respective quantities.    

The key conclusions are the following.  In a competitive adsorption process, 
divalent ions displace the condensed monovalent ions at modest temperatures 
and for reasonable 
values of the dielectric mismatch parameter. At similar physical conditions,
increasing divalent salt concentration can induce complete neutralization
and resulting contraction of the polyelectrolyte chain. With further increase
of divalent salt in the solution, condensed divalent counterions overcharge and 
reexpand  chains for significantly large ranges 
of physical parameters $\delta$ and $l_B$. A substantial
fraction of negative salt coions (Cl$^-$) also condenses on the 
monomer-divalent ion-pair which reduces
the degree of overcharging but does not eliminate it. However, for modest 
temperatures in a generic solvent (water at room temperature), there is always a
critical degree of dielectric heterogeneity below which neither chain collapse nor 
charge reversal occurs irrespective of the amount of divalent salt in solution. 
Further, the dependencies of the regularization of charge and size of the polymer 
on salt concentration, temperature, and the dielectric
mismatch parameter are relatively robust functions of $l_B$, $\kappa^{-1}$, and 
$\delta$, not depending sensitively on the microscopic details of the 
charge complexes. For a moderate
presence of salt, both the degree of ionization and chain size 
are typically smaller for divalent salts than for
monovalent salts. A higher dielectric mismatch and a lower temperature
enhance condensation of all types of counterions and coions resulting in
the achievement of isoelectric point at lower divalent salt concentrations. 
A typical state diagram for polyelectrolyte charge is predicted 
(Fig. \ref{cartoon}) in which below a 
critical Coulomb strength (proportional to $1/T, 1/\epsilon$ or $\delta$)
there is no overcharging with increasing divalent salt. Above this critical 
strength, electrostatics is strong enough to condense sufficient number of 
counterions inducing overcharging. For higher salt concentrations, we expect
recharging of the polyelectrolyte due to either screening of 
electrostatics\cite{praetal-04} 
(low Coulomb strength) or negative coion (Cl$^-$) condensation (high Coulomb 
strength). The dashed line indicates that in this regime of salt concentration
the theory only predicts qualitative results. If ion-bridging is present, a
simple theory based on this model predicts first order chain collapse.
Our theory shows that overcharging is an outcome of both correlation-induced
adsorption related to the discreteness of divalent cations and dielectric
heterogeneity related to the local chemical structure of polyelectrolytes.
  
The rest of the article is organized as follows: in Sec.II, we develop the 
theory and in Sec.III, we present the results discussing competitive
adsorption (III.A), chain collapse (III.B), overcharging (III.C), coion
condensation (III.D), free energy (III.E), state diagrams for various regimes
of polymer charge (III.F), and the
bridging scenario (III.G). Conclusions are summarized in Sec.IV.

\section{Theory}

\setcounter{section}{2}
\setcounter{equation}{0}

Following Muthukumar's work with monovalent 
salt\cite{mut-04}, we consider a linear flexible polyelectrolyte chain of $N$
monomers in a solution of volume $\Omega$, with the center of mass of the chain
at the origin of the coordinate system. Each monomer is monovalently charged
(negative) and of length $l$.  There can be either or both monovalent and 
divalent 
salts in the solution (say, water), which being electroneutral at all times 
will have a maximum of $N$
monovalent counterions in addition to the salt ions. We assume that the 
counterion from the monovalent salt (say, Na$^+$ from NaCl) is chemically 
identical to the counterion from the polymer (say, Na$^+$ from NaPSS).
Similarly, the coions from both types of salts  
are of the same species (say, Cl$^-$ from NaCl and BaCl$_2$). At 
any time, both monovalent and divalent counterions (say, Ba$^{++}$
from BaCl$_2$ as divalent counterions) can condense on separate monomers. In 
addition,
the Ba$^{++}$-monomer ion pair is viewed as a positive monovalent ion, and
the negative coions (Cl$^-$) will condense on some of these pairs as 
counterions. Therefore, if $M_1$ monovalent counterions and $M_2$ divalent
counterions get adsorbed on the chain ($M_1 + M_2 \le N$), and $M_3$
(negative) coions condense on Ba$^{++}$-monomer ion pairs ($M_3 \le M_2$), the
effective (or average) degree of ionization of the entire chain is 
$f = (1 - M_1 - 2 M_2 + M_3)/N$.  $R_g$ is the radius of gyration of the chain.
$c_{s1}$ and $c_{s2}$ are, respectively, the number concentrations
of the added monovalent and divalent salts. Both types of salts are fully 
dissociated into $n_1$ monovalent counterions (Na$^+$), $n_2$ divalent 
counterions (Ba$^{++}$) and $n_1 + 2 n_2$ negative coions (Cl$^-$). Therefore,
$c_{s1} = n_{1+}/\Omega$ and $c_{s2} = n_{2+}/\Omega$. The free energy of
the system, consisting of the chain, condensed and mobile counterions, and the
solution, would depend on four independent variables: $M_1, M_2, M_3$, and 
$R_g$.  The theory\cite{mut-04} aims to evaluate
$M_1, M_2, M_3$, and $R_g$ self-consistently by calculating the free energy
$F$ of the system as a function of all these variables and electrostatic
parameters. The equilibrium values of the four major variables are 
obtained by minimizing $F$ simulatnously with respect to these variables. 
Therefore, this requires an extension of the previous theory of minimization 
with two variables to a theory with four variables.

As before\cite{mut-04}, the free energy $F$ has six contributions $F_1, F_2,
F_3, F_4, F_5$, and $F_6$ related, respectively, to (i) entropy of mobility 
of the condensed
counterions and coions on the polymer backbone, (ii) translational entropy of
uncondensed counterions and coions (including salt ions) which are mobile 
within the volume $\Omega$, (iii) electrostatic fluctuation interaction 
(Debye-H\"{u}ckel) among all unadsorbed counterions and coions  except the polymer, 
(iv) the
unscreened electrostatic (Coulomb) energy of monomer-counterion pairs (both
monovalent and divalent counterions) and monomer-counterion-coion triplets,
(v) free energy of the polyelectrolyte with an average degree of ionization
$f$, and (vi) electrostatic correlation involving the neutral ion-pairs or 
ion-triplets along the backbone of the polymer.

\subsection*{A. The free energy}

To determine the entropic contribution from the condensed counterion and 
coions, we note that there are $N$ monomers, $M_1$ condensed monovalent 
counterions (Na$^+$), $M_2 - M_3$ condensed divalent counterions (Ba$^{++}$) 
with no negative coion (Cl$^-$') condensation, and $M_3$ 
ion-triplets ('monomer-Ba$^{++}$-Cl$^-$'). Therefore, $N - M_1 - M_2$ monomers
remain bare charged with no ions condensed on them. Consequently, the partition 
function is
\bea
\label{f1zdivcl}
Z_1 = \frac{N!}{(N - M_1 - M_2)! M_1! (M_2 - M_3)! M_3!}.
\eea
We define, 
\bea
\label{alphas}
\alpha_1=\frac{M_1}{N}; \qquad \alpha_2=\frac{M_2}{N}; \qquad \alpha_3
=\frac{M_3}{N}
\eea
Using $F_1 = - K_B T \ln Z_1$, we have,
\bea
\label{f1final}
\frac{F_1}{N k_B T} &=& (1 - \alpha_1 - \alpha_2)\log (1 - \alpha_1 - \alpha_2)
    + \alpha_1\log \alpha_1 \\ \nonumber 
    &+& (\alpha_2 - \alpha_3)\log (\alpha_2 - \alpha_3) + \alpha_3\log \alpha_3. 
\eea
The above expression implies two obvious constraints:
\bea
\label{constraints1}
\alpha_1 + \alpha_2 \le 1 \qquad \mbox{and} \qquad \alpha_3 \le \alpha_2.
\eea

To determine the translational entropy of the uncondensed ions which are 
distributed in the bulk volume $\Omega$, we count as mobile ions:
$N - M_1 + n_{1+}$ monovalent counterions (Na$^+$), $n_{2+} - M_2$ divalent 
counterions (Ba$^{++}$), and $n_{1+} + 2 n_{2+} - M_3$ monovalent negative 
coions (Cl$^-$).  Therefore, the partition function related to the translational 
free energy in volume $\Omega$ would be
\bea
\label{f2zdivcl}
Z_2 &=& \frac{\Omega^{N - M_1 + n_{1+} + n_{2+} - M_2 + n_{1+} + 2 n_{2+} - M_3}}
    {(N - M_1 + n_{1+})! (n_{2+} - M_2)! (n_{1+} + 2 n_{2+} - M_3)!}\\\nonumber
    &=& \frac{\Omega^{N - M_1 - M_2 - M_3 + 2 n_{1+} + 3 n_{2+}}}
      {(N - M_1 + n_{1+})! (n_{2+} - M_2)! (n_{1+} + 2 n_{2+} - M_3)!}.
\eea
Relating numbers of ions and their concentrations is helpful. We note
\bea
\label{css}
N = \rho \Omega; \qquad n_{1+} = \frac{c_{s1} N}{\rho}; \qquad n_{2+} 
= \frac{c_{s2} N}{\rho}.
\eea
With use of $F_2 = - K_B T \ln Z_2$ and after some calculations we arrive at
\bea
\label{f2final}
\frac{F_2}{N k_B T} &=& (1 - \alpha_1 + \frac{c_{s1}}{\rho})
\log (\rho(1 - \alpha_1) + c_{s1})
   + (\frac{c_{s2}}{\rho} - \alpha_2)\log (c_{s2} - \rho \alpha_2) \\ \nonumber
  &+& (\frac{c_{s1}}{\rho} + 2 \frac{c_{s2}}{\rho} - \alpha_3)
\log (c_{s1} + 2 c_{s2} - \rho \alpha_3) \\ \nonumber
    &-& \{(1 - \alpha_1 - \alpha_2 - \alpha_3) 
+ 2 \frac{c_{s1}}{\rho} + 3 \frac{c_{s2}}{\rho}\}. 
\eea
Here, the constraint would be 
\bea
\label{constraints2}
M_2 \le n_{2+}.
\eea

The free energy contribution from the correlations of all dissociated ions can 
be calculated in the electrostatic free energy
\bea
\label{f3initial1}
F_3 = - k_B T \frac{\Omega \kappa^3}{12 \pi},
\eea
where the inverse Debye length $\kappa$ is given by
\bea
\label{kappasquare1}
\kappa^2 = 4 \pi l_B \sum_i Z_i^2 n_i/\Omega.
\eea
This result is obtained from the Debye-H\"{u}ckel (DH) theory and caution must be 
exercised to identify regimes where the DH theory is at best a gross
approximation. Here, $Z_i$ is the valency of the dissociated ion of the $i$-th 
species. In this case [see the text before Eq. (\ref{f2zdivcl})], 
\bea
\label{kappasquare2}
\kappa^2 &=& 4 \pi l_B \{N - M_1 + n_{1+} + 4(n_{2+} - M_2) + n_{1+} + 2 n_{2+} 
                                                    - M_3\}/\Omega \\ \nonumber
         &=& 4 \pi l_B (N - M_1 - 4 M_2 - M_3 + 2 n_{1+} 
                                                + 6 n_{2+})/\Omega \\ \nonumber
         &=& 4 \pi l_B \left\{\rho(1 - \alpha_1 - 4 \alpha_2 - \alpha_3) 
                                              + 2 c_{s1} + 6 c_{s2} \right\}.
\eea
Using the definitions from Eq. (\ref{css}) we reach
\bea
\label{f3final}
\frac{F_3}{N k_B T} = -\frac{1}{3}\sqrt{4 \pi} l_B^{3/2} \frac{1}{\rho} 
\{\rho (1 - \alpha_1 - 4 \alpha_2 - \alpha_3) + 2 c_{s1} + 6 c_{s2}\}^{3/2}.
\eea

To determine the electrostatic energy gain due to condensation of all sorts of
ions (Na$^+$, Ba$^{++}$, and Cl$^-$), we recount different ion pairs 
and triplets which form after condensation. On the polymer chain, there are
$\alpha_1 N$ pairs of 'monomer(-1) and Na$^+$ ion',
$(\alpha_2 - \alpha_3) N$ pairs of 'monomer(-1) and Ba$^{++}$ ion', and 
$\alpha_3 N$ triplets of 'monomer(-1), Ba$^{++}$ and Cl$^-$ ions'. In addition, 
$(1 - \alpha_1 - \alpha_2) N$ monomers(-1) remain charge uncompensated. 
In Muthukumar's theory\cite{mut-04}, the dielectric mismatch parameter 
$\delta$ was conceived
to address a local dielectric constant $\epsilon_l$ in the vicinity of the chain
backbone. Experiments have shown\cite{meheic-84} $\epsilon_l$ to be almost an 
order of magnitude less than its bulk value $\epsilon$ (around 78 in water)
near polyelectrolyte or protein backbone. The dielectric constant increases 
exponentially\cite{lampac-97,roublo-98} from the fractional to its full bulk
value over a distance of 1-10$\AA$ from the chain monomers. 
$\delta = (\epsilon l/\epsilon_l d)$ was introduced\cite{mut-04}, where $d$ is
the dipole length of the monomer-monovalent counterion ion-pair. A glance at 
Fig. \ref{schematic} reveals that $\delta$ in above form only applies to the 
monomer-monovalent (Na$^+$) and monomer-divalent (Ba$^{++}$) ion pairs, but 
not to the divalent counterion-monovalent coion (Ba$^{++}$-Cl$^-$) ion pair
in the monomer-divalent counterion-monovalent coion triplet. In an ion pair,
there are two ions involved with a fixed distance between them. For the 
triplet, however, there are three lengths involved (for example, 
Ba$^{++}$-monomer, Ba$^{++}$-Cl$^-$ and Cl$^-$-monomer), and interpretation
of $\delta$ is a bit tricky. We introduce a parameter $\delta_2$ for the
'monomer-Ba$^{++}$-Cl$^-$' triplet. $\delta_2$ is expected to be less than 
$4 \delta$ (the value it would have assumed if there were two point charges,
$+2e$ and $-2e$, respectively), but the determination of its actual value would 
probably require a microscopic treatment. In principle,  $\delta_2$ would be 
a function of $\delta$. For simplicity, we assume all ions and monomers to be
of the same size, and determine $\delta_2$ in what follows. 

First of all, we write the electrostatic energy of condensation in terms of
$\delta$ and $\delta_2$:
\bea
\label{f4final}
\frac{F_4}{N k_B T} = - \alpha_1 \delta \tilde{l_B} 
- 2 (\alpha_2 - \alpha_3) \delta \tilde{l_B} - \alpha_3 \delta_2 \tilde{l_B}, 
\eea
where $\tilde{l_B} = l_B/l$ and the terms containing $\delta$ are written
following Ref. 1 (with $l \sim d$).   
Note that if one assumes that the local dielectric constant $\epsilon_l$ 
applies only to the Ba$^{++}$-monomer pair, but
not to the Ba$^{++}$-Cl$^-$ pair ($\epsilon = \epsilon_{water}$ in that case), 
then $\delta_2$ turns out to be
\bea
\label{delta2low}
\delta_2 = \left( 2 + \frac{2}{\delta} \right) \delta.
\eea
Those should be the lowest values for $\delta_2$. On the other hand, if 
$\epsilon_l$ applies to both Ba$^{++}$-monomer and
Ba$^{++}$-Cl$^-$ pairs, then 
\bea
\label{delta2high}
\delta_2 = 4 \delta.
\eea 
Those should be the highest values for $\delta_2$. In practice, $\delta_2$ 
would be somewhere between these two limiting sets of values. 
We choose the dielectric constant to be $\epsilon_l$ for the Ba$^{++}$-monomer 
pair and $(\epsilon_l+\epsilon_{bulk})/2$ for the Ba$^{++}$-Cl$^-$ pair. 
Then, $\delta_2$ turns out to be
\bea
\label{delta2used}
\delta_2 = \left( 2 + \frac{4}{\delta+1} \right) \delta.
\eea
Point worthy of note is that the repulsion between the monomer and the Cl$^-$ 
ion has been ignored; it would bring a very small correction in all three cases
above. Although the counterion distribution and chain conformations are 
sensitively dependent on $\delta_2$, we later show that the very basic 
qualitative results do not change if $\delta_2$ is assigned any value 
in the range mentioned above. Therefore, we would use Eq. (\ref{delta2used})
in all of our representative calculations, unless mentioned otherwise. 

The free energy of the flexible polyelectrolyte chain is obtained by the 
variational method\cite{mut-04,mut-87} in which one starts from the Edwards
Hamiltonian,
\bea
\label{edwards}
H &=& \frac{3}{2l}\int_0^L ds \left( \frac{\partial {\mathcal R}(s)}
{\partial s}\right)^2 + \frac{w}{2} \int_0^L ds \int_0^L ds' \delta 
\left({\mathcal R}(s) - {\mathcal R}(s')\right) \\ \nonumber
&+& \frac{l_B}{2} \int_0^L ds 
\int_0^L ds' \frac{1}{|{\mathcal R}(s) - {\mathcal R}(s')|} \exp-\kappa
|{\mathcal R}(s) - {\mathcal R}(s')|,
\eea  
where $L=Nl$, ${\mathcal R}(s)$ is the position vector of the chain at arc 
length $s$, and $w$ is the strenth parameter for all short-ranged hydrophobic
or excluded volume effects. An effective expansion factor $l_1$ is defined
as follows:
\bea
\label{expansion}
\langle R^2 \rangle = N l l_1 \equiv N l^2 \tilde{l_1} = 6 R_g^2.
\eea
Here, $\langle R^2 \rangle$ is the mean square end-to-end distance, and $l_1$
effectively measures the swelling of the chain compared to a Gaussian chain.
Assuming uniform, spherically symmertic  expansion or contraction of the chain,
and by extremizing the free energy we obtain
\bea
\label{f5final}
\frac{F_5}{N k_B T} &=& \frac{3}{2 N}\left( \tilde{l_1} - 1 
- \log \tilde{l_1} \right) + \frac{4}{3}\left( \frac{3}{2 \pi} \right)^{3/2}
\frac{w}{\sqrt{N}} \frac{1}{\tilde{l_1}^{3/2}} \\ \nonumber
&+& 2 \sqrt{\frac{6}{\pi}} f^2
\tilde{l_B} \frac{N^{1/2}}{\tilde{l_1}^{1/2}} \Theta_0 (a),
\eea
where 
\bea
\label{theta}
\Theta_0 (a) = \frac{\sqrt{\pi}}{2}\left( \frac{2}{a^{5/2}}
- \frac{1}{a^{3/2}} \right) \exp (a) \mbox{erfc} (\sqrt{a}) + \frac{1}{2 a}
+ \frac{2}{a^2} - \frac{\sqrt{\pi}}{a^{5/2}} - \frac{\sqrt{\pi}}{2 a^{3/2}},
\eea
where
\bea
\label{a}
a \equiv \tilde{\kappa}^2 N \tilde{l_1}^2 / 6.
\eea
Here, $\tilde{\kappa} = \kappa l$. We further define two more dimensionless
variables, $\tilde{\rho}= \rho l^3$ and $\tilde{c_{si}} = c_{si} l^3$, where
$i$ stands for the ion species. The
important factor $f$ is our previously defined average degree of ionization
and is given by 
\bea
\label{doi}
f = 1 - \alpha_1 - 2 \alpha_2 +  \alpha_3.
\eea
The justification of using the variational result as the polymer free energy
has been discussed in the previous paper\cite{mut-04}. We assume that similar
arguments are still valid for this work. Any other alternative function of
average degree of ionization ($f$) and radius of gyration ($R_g$) for $F_5$ may
be used in place of Eq. (\ref{f5final}). 

Till now we have considered the electrostatic interaction only between monomers
the effective charge of which, with or without condensed ions, is
non-zero. In other words, the third-term in the polymer free energy $F_5$
[Eq. (\ref{f5final})] addresses the electrostatic interaction (within
Debye-H\"{u}ckel approximation) between monomers with non-zero effective 
monopole charges. Further,
$F_5$ considers only the monopole contribution of each ion-pair or ion-triplet. 
For example,
a monomer-Na$^+$ pair and a monomer-Ba$^{++}$-Cl$^-$ triplet would be treated
equally by $F_5$, although they have quite different electrostatic effects. 
Similarly, a monomer-Ba$^{++}$ pair would be simply treated as a +1 charge,
although the pair will have additional dipole effects. These additional
dipole or higher order multipole effects would be critical when the 
average charge of the chain is close to zero. It has been 
shown\cite{winetal-98,liumut-02,mut-04}
that these ion-pair effects play a key role to collapse a chain in
presence of monovalent counterions at very low temperatures (i.e, when the
degree of ionization is negligible). In the previous paper\cite{mut-04},
this correlation among neutral ion-pairs and bewteen neutral ion-pairs and 
charged monomers were addressed by short-ranged $\delta$-function potentials
which led to free energy contribution of the form
\bea
\label{f6final}
\frac{F_6}{N k_B T} \sim \frac{4}{3}\left( \frac{3}{2 \pi} \right)^{3/2}
w_i \delta^2 \tilde{l_B}^2 \frac{1}{\sqrt{N}} \frac{1}{\tilde{l_1}^{3/2}},
\eea
where $w_i$'s ($< 0$) are temperature dependent parameters, and are 
different for dipole-dipole and dipole-monopole interactions. These 
contributions are attractive and would modify the excluded volume
interaction [the second term in $F_5$, Eq. (\ref{f5final})].
They can significantly reduce the size of the chain only around the 
isoelectric point ($f \sim 0$), and the type of collapse is generally
continuous or second order. 

In addition to the short-ranged dipole correlations, there can be long-ranged
attraction  between monomers mediated by multivalent 
counterions\cite{roublo-96,haliu-97}. This 
attractive correlation between counterions may compensate 
the residual Coulomb repulsion of the chain even at higher degrees of 
ionization\cite{goletal-99}, and the extended conformation of the chain may
become unstable. This can as well be treated with the concept of ion 
'bridging'\cite{groetal-02,liuetal-03}. It is still not conclusively known 
what kind of collapse this correlation-induced long-ranged attraction 
may induce. We leave out the short-ranged correlation effects near the 
isoelectric point in our present analysis. In Section III.G, we give 
preliminary results of an ion bridging theory leading to global instability of a
polyelectrolyte chain based on our model. The point worthy of note here is that 
the bridging interaction 
reduces\cite{liuetal-03} the effective value of the excluded volume parameter $w$. 
Therefore, for higher values of $w$, only very high Coulomb strength or divalent 
salt concentration will allow the bridging effect to take place. In most of our
analysis (only except Section III.G), we assume $w$ to be high enough to render 
the bridging effect to be negligible. Although we assume $w$ to be zero except 
in Section III.G, that will imply the 'no-bridging' scenario in which choosing 
non-zero positive value of $w$ only brings minor quantitative changes to our 
results. When bridging is included (Section III.G), however, $w$ is a very 
important parameter affecting the transition salt concentration or Coulomb 
strength. A more detailed analysis of the role of multivalent cations in 
collapsing a polyelectrolyte and the related order of the transition will 
be presented in a future publication.

\section{Results and Discussion}

\setcounter{section}{3}
\setcounter{equation}{0}

We can express the total free energy $F = F_1 + F_2 + F_3 + F_4 + F_5$
in terms of the fraction of condensed counterions and coions ($\alpha$'s),
size of the polymer ($l_1$), temperature and bulk dielectric constant ($l_B$), 
degree of polymerization ($N$),
monomer density ($\rho$), monovalent and divalent salt concentrations ($c_s$'s),
and local dielectric mismatch parameters ($\delta$ and $\delta_2$). 
The goal is to self-consistently
determine the fractions of condensed ions ($\alpha_1, \alpha_2$, and $\alpha_3$)
and the size ($R_g = \sqrt{(N l l_1/6)}$) that minimize the free energy. It is
a simultanous minimization with respect to four variables ($\alpha_1, \alpha_2, 
\alpha_3, \tilde{l_1}$) instead of two in the previous paper\cite{mut-04}, and 
it is best performed numerically. Compared to a neutral system, there are two 
additional length scales in a charged system. They are the Bjerrum length ($l_B$)
related to the Coulomb interaction and the Debye length ($\kappa^{-1}$) 
introduced by screening due to dissociated ions including salt ions. This 
formalism invokes a third length scale due to the dielectric mismatch parameter 
$\delta$. Therefore, the important parameters on which we base our analysis
are $l_B$, the salt concentrations ($c_{s1}$ and $c_{s2}$),
and $\delta$.

\subsection*{A. Competitive adsorption}

	We start with an isolated polyelectrolyte at low concentrations and at 
modest temperatures. In the first observation, the concentration of the divalent 
salt ($c_{s2}$) is increased from while keeping the concentration of the 
monovalent salt at zero ($c_{s1} = 0$). We have chosen a higher and a lower
value of $\delta=2.5$ and 1.5, respectively. Generally at higher values of 
$\delta$, fractions of condensed ion species are expected to increase. We notice 
that [Fig. \ref{delta-1p5-2p5-alphas-l1t}] both divalent counterions and 
negative monovalent coions condense progressively in higher numbers 
($\alpha_2$ and $\alpha_3$, respectively) with increasing divalent salt
concentration. The number of monovalent counterions ($\alpha_1$), however, 
decreases with increasing $c_{s2}$. This implies that in this competitive
adsorption process, condensed monovalent counterions, when challenged by a 
divalent salt, are replaced by divalent counterions. This happens for the entire
physical range of the dielectric mismatch paremeter (we will later show 
in the diagrams of charged states). The variable values  chosen in this specific 
calculation
are degree of polymerization $N = 1000$ and monomer density $\tilde{\rho} 
= \rho l^3 = 0.0005$ at  $\tilde{l_B} = 3.0$ (value related to flexible polymers 
of the type sodium polystyrene sulphonate (NAPSS) in water at room temperature).  
$\delta_2$ is given by Eq. (\ref{delta2used}) throughout the paper, unless 
noted otherwise. For $\delta=1.5$, only $5\%$ of monomers are neutralized 
by monovalnt counterions at no salt situation ($c_{s1}, c_{s2} = 0$) 
[Fig. \ref{delta-1p5-2p5-alphas-l1t}(a)] whereas 
the number increases to $35\%$ for $\delta=2.5$ 
[Fig. \ref{delta-1p5-2p5-alphas-l1t}(b)]. 
At this higher $\delta$ value, almost all available divalent counterions 
condense displacing the monovalent counterions with increasing $c_{s2}$. 
Negative monovalent coions (Cl$^-$) also condense on the 
monomer-divalent ion-pair substantially. $f$ decreases monotonically and 
{\it reverses sign} when $c_{s2}$ is still well below the monomer concentration
$\rho$. At $c_{s2} \sim 80\%$ of $\rho$, $\alpha_1$ drops below $5\%$ and 
$\alpha_2$ is about $80\%$ implying that almost all available divalent 
counterions have condensed. The size of the chain decreases steeply 
[Fig. \ref{delta-1p5-2p5-alphas-l1t}(d)]
due to rapid neutralization, but increases beyond the isoelectric
point ($f=0$) due to repulsion among divalent cations that overcharge the 
chain.  The size is dictated by the third term in Eq. (\ref{f5final}) at these 
salt concentrations. For higher values of $c_{s2}$, more negative coions condense 
on the chain to reduce the (over)charge of the chain marginally. Number of 
condensed monovalent ions, however, decrease to zero monotonically.  
	
    For the lower value of $\delta$, the original sign of the polyelectrolyte 
charge ($f$) is preserved even at higher divalent salts 
[Fig. \ref{delta-1p5-2p5-alphas-l1t}(a)] with the minimum absolute
degree of ionization  being around 0.27. Consequently, the size of the chain
[Fig. \ref{delta-1p5-2p5-alphas-l1t}(c)] remains substantially bigger than 
the Gaussian value for the entire range of salt cincentration. 
        
\subsection*{B. The chain collapse}

	In Fig. \ref{delta-1p5-2p5-alphas-l1t}(b), we have noticed
that the polyelectrolyte net charge due to condensation of divalent counterions
becomes negligible as soon as the salt concentration reaches
half the polymer concentration ($c_{s2} \sim \rho/2$). Consequently,
the chain collapses to its Gaussian size at around this isoelectric point. To
compare with the case of monovalent counterions, we plot both the degree of 
ionization $f$ and the expansion factor $\tilde{l_1}$ at $\tilde{l_B} = 3.0$ 
in Fig. \ref{delta-2p5-cs1-cs2}. 
The other parameters are: $N = 1000, \delta = 2.5$, and 
$\tilde{\rho} = 0.0005$. We notice that for the monovalent salt, degree of
ionization $f$ of the polyelectrolyte decreases moderately and monotonically, 
and never changes sign.
Consequently, the size ($l_1$ or $R_g$) also decreases monotonically with the 
Gaussian statistics being obtained only at very high salt concentrations (or
at very low temperatures). For the divalent salt, however, the isoelectric 
point is achieved as soon as there are sufficient number of divalent 
counterions available to neutralize the chain. That happens at a very 
low $c_{s2}$. As a result, the polyelectrolyte collapses (Gaussian statistics)
near this isoelectric point. 

	This collapse of a generic polyelectrolyte (NAPSS) in water occurs 
for modest values of $\delta$, at a modest presence of the divalent salt, 
and at room temperature. This 
phenomenon has long been noticed theoretically\cite{sololv-00}, and very 
recently in experiments\cite{zhaetal-01,muretal-03,besetal-07} and 
simulations\cite{hsilui-06}. 

\subsection*{C. The issue of overcharging}

        Our theory predicts that charge neutralization and subsequent charge
reversal would occur to an isolated flexible polyelectrolyte in 
aqueous solutions at room temperature and at a modest presence of a 
divalent salt. The parameter $\delta$ in our theory
plays an important role in the charge reversal induced by counterion condensation. 
Temperature is also an important factor regulating the relative weight of 
electrostatic energy gain of ion condensation. To show these effects,
we plot the degree of ionization $f$ and the expansion factor $\tilde{l_1}$ 
of the chain as functions of Bjerrum length $l_B$ (inverse temperature and bulk
dielectric constant) for 
various $\delta$ values in Fig. \ref{cs2-delta-lb}. The other parameters  are:
$N=1000, \tilde{\rho} = 0.0005, \tilde{c_{s1}} = 0$ and $\tilde{c_{s2}} 
= 0.0005 = \tilde{\rho}$.
The concentrations of the divalent salt and the polymer is chosen to be equal  
to ensure the availability of enough divalent
ions to condense over every monomer if physical conditions permit. 
In Fig. \ref{cs2-delta-lb}(a), we notice
that there is negligible condensation for $\delta = 1$ (which is the 
comparable value of $\delta$ in simulations\cite{liuetal-03,hsilui-06}). Similar
to the monovalent case [Fig. 2(a) and 3 in Ref.\cite{mut-04}], the chain is 
neutralized only at very low temperatures (there is a factor of two in $l_B$ 
because the Coulomb energy gain for each ion-pair is twofold for divalent ions).
At no temperature there is overcharging for $\delta=1$. There is, however, a 
drastic
qualitative change in the dependencies of $f$ and $R_g$ on $l_B$ for $\delta$
values 2 and above. At a particular temperature $T_0$, the chain is
neutralized and if $T$ is further reduced, overcharging occurs (and the
chain swells). $T_0$ is higher for higher values of $\delta$ as expected
($T^{-1}$ and $\delta$, both favor higher degree of condensation). The absolute 
value of maximum overcharge and re-swelled size increase with $\delta$ as well.
In particular, the reswelled size is larger than the original swelling 
for $\delta=4.0$. This is despite the absolute effective charge being lower
at the maximal $reswelling$ because, at this point, the Coulomb strength for this 
large $\delta$ value is high enough to have repulsion between monomers stronger 
than at point of maximal $swelling$ as we increase $\tilde{l_B}$.
Another point of note is that for higher temperatures, just as for 
monovalent counterions, only a fraction of available divalent ions condense.
The optimal temperature at which the chain reexpansion is maximum shifts
to a higher value with higher values of $\delta$. It might be instructive
to note that we increase $c_{s2}$ in Fig. \ref{delta-1p5-2p5-alphas-l1t}(b), 
by fixing the system at the abscissa value of 3 in Fig. \ref{cs2-delta-lb} 
and staying on the $\delta = 2.5$ curve.  For very low temperatures, sufficient 
number of negative coions (Cl$^-$) condense to gradually re-neutralize the 
chain for all $\delta$ values.   

	To further explore the issue of overcharging, we plot $f$ and 
$\tilde{l_1}$ against $\delta$ for $\tilde{l_B} = 3$ 
in Fig. \ref{cs2-delta}. The other parameters are the same as in 
Fig. \ref{cs2-delta-lb}. At room temperature in aqueous solutions, there
can be no overcharging unless $\delta > 1.7$. Only for $\delta$ as high
as 1.7, the dieletric heterogeneity would be strong enough to electrostatically
guide enough divalent ions that condense and reverse the charge of the
chain. The strong sensitivity of the total charge and conformation of the 
polymer on $\delta$ is manifest in Fig. \ref{cs2-delta}, in which $f$ decreases 
from about 93$\%$ to zero (and subsequently $\tilde{l_1}$ decreases from
about 25 to 1 - the Gaussian value) for $\delta$ changing only from
1-1.7. For very high values of $\delta$, Cl$^-$ ions condense progressively 
at higher numbers to reduce overcharging.

\subsection*{D. Condensation of Cl$^-$ ions}

In a monomer-Ba$^{++}$-Cl$^-$ ion-triplet, the colinear arrangement of
the three charges, in the same order as written here, is 
electrostatically the most favorable one. This is true regardless of the strength 
of $\delta$ or the range of the local dielectric constant $\epsilon_l$. 
We assume that all triplets in the system have this specific colinear 
arrangement in which the line joining the charges is perpendicular to the chain 
backbone [see Fig. \ref{schematic}]. We have discussed before the
ambiguity in determining the electrostatic energy gain per triplet formation.
The highest and lowest values permitted by physical conditions of the strength 
paramater $\delta_2$ related
to the formation of the triplet has been determined in terms of $\delta$ 
through Eqs. (\ref{delta2low})-(\ref{delta2high}). Although suggested by a few
authors\cite{roublo-98}, we did not consider a different local dielectric 
constant for isolated Ba$^{++}$ or Cl$^-$ ions.
The dielectric constant relevant for the Coulomb interaction between the 
Ba$^{++}$ and Cl$^-$ ions in Eqn. (\ref{delta2used}) is different than $\epsilon$. 
This is because of the fact that these two ions are in the local environment of the 
{\it chain backbone}, and consequently, the local dielectic behavior of the 
polyelectrolyte (i.e., $\delta$) exclusively determines the value of $\delta_2$.
In this theory, to minimize the number of adjustable parameters, 
we assume that the sizes of the ions (Ba$^{++}$, Cl$^-$, and Na$^+$) 
are of the order of the size $l$ of the monomer. To illustrate that these 
approximations
do not compromise the generality of the problem, we plot the fraction
of condensed ions ($\alpha_1, \alpha_2, \alpha_3$), degree of ionization ($f$),
and the size expansion factor ($\tilde{l_1}$) of the polymer as functions of 
$\delta_2$. We
vary $\delta_2$ from its lowest [Eq. (\ref{delta2low})] to its highest 
value [Eq. (\ref{delta2high})] for a specific value of $\delta$. In
Fig. \ref{delta2-ovch-noovch-alphas-l1t}(a), we choose $\delta = 2.5$, 
for which overcharging is evident for $\tilde{l_B} = 3.0$. Other parameters are:
$N = 1000, \tilde{\rho} = 0.0005 = \tilde{c_{s2}}$ and $\tilde{c_{s1}} = 0$.
We notice that with increasing $\delta_2$, progressively larger number of
Cl$^-$ ions condense on the chain. This reduces overcharging and close to the
highest value of $\delta_2$, overcharging would be marginally eliminated.
This result has close resemblance to recent simulations\cite{hsilui-06},
where it is observed that smaller ion sizes (which leads to 
effectively higher $\delta_2$ values) reduce the degree of overcharging.
In Fig. \ref{delta2-ovch-noovch-alphas-l1t}(b), we choose $\delta = 1.5$, 
for which there is no overcharging at $\tilde{l_B} = 3.0$. We notice 
that a change in $\delta_2$
has negligible effect on both $f$ and $R_g$ in this case. However, 
fractions of both condensed divalent counterions and monovalent coions 
($\alpha_2$ and $\alpha_3$, respectively) increase with $\delta_2$, keeping
the overall degree of ionization ($f$) approximately unaltered.  

\subsection*{E. The free energy profile}

One of the advantages of our equilibrium adsorption theory is that it is 
possible to compare the contributions of different factors
in the total free energy ($F_1$ to $F_5$) as functions of the critical 
parameters. The major
conclusion of the theory\cite{mut-04} has been that the equilibrium
distribution of counterions and the size of the polyelectrolyte are determined
essentially by the competition between the translational entropy of dissociated
ions and the Coulomb energy gain of condensed ions. This is indeed borne out
by our calculation in the presence of divalent salts too, as shown in 
Fig. \ref{free-energy}. In Fig. \ref{free-energy}(a), the separate parts
of the free energy are plotted against the Bjerrum length $\tilde{l_B}$
for a fixed divalent salt concentration ($\tilde{c_{s2}} = 0.0005$, equal
to the monomer density) and for a specific strength of dielectric mismatch
($\delta = 2.5$). The major contributions to the total free energy come
from the translational entropy $F_2$ [Eq. (\ref{f2final})] and the Coulomb 
free energy $F_4$ [Eq. (\ref{f4final})]. For higher temperatures (lower $l_B$'s),
the entropic term is favored as electrostatics remains negligible compared
to thermal fluctuations. For lower temperatures,
electrostatics becomes progressively relevant, and many ions condense 
reflecting substantial gains in $F_4$. The entropic contribution 
$F_1$ [Eq. (\ref{f1final})] related to
the mobility of condensed ions along the backbone has negligible effect,
and so does the Debye-H\"{u}ckel contribution $F_3$ [Eq. (\ref{f3final})] at these
salt concentrations. 

In Fig. \ref{free-energy}(b), similar free energy components are plotted 
against $\delta$ at the same salt concentration and for $\tilde{l_B} = 3.0$. 
The curves in (a) and (b) are remarkably similar 
demonstrating the equivalence of the parameters $l_B$ and $\delta$. 
According to this adsorption theory, reduction of any of temperature, the
bulk dielectric constant $\epsilon$ or the local dielectric constant 
$\epsilon_l$ (near the hydrophobic regions of the
chain backbone) by a similar factor would induce very similar effects to 
polyelectrolyte behaviors. This is especially valid for modest values of
$l_B$ and $\delta$.  

\subsection*{F. The diagrams of charged states}

\begin{figure}[p] \centering
\hspace*{0cm}{\epsfxsize= 16cm \epsfbox{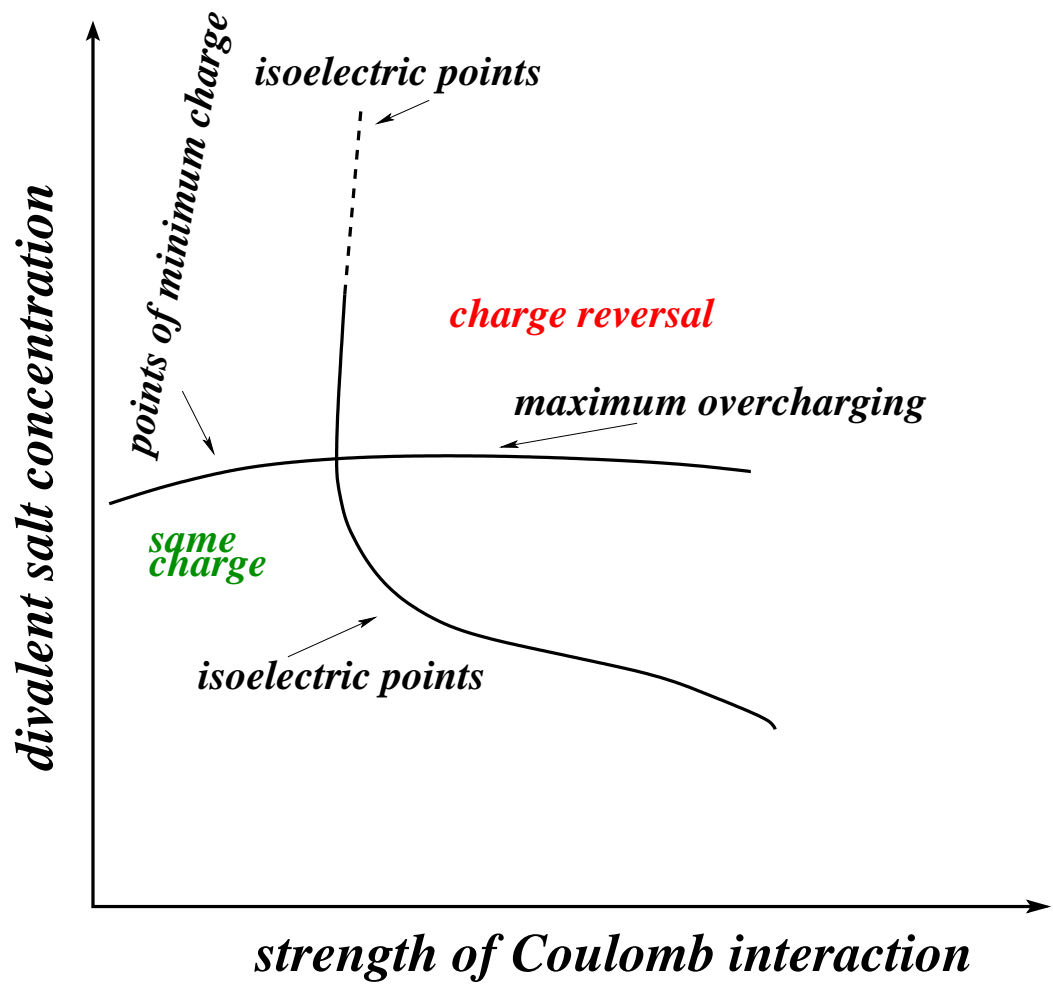}}
\bigskip
\vspace*{0.0cm}
\vspace*{0cm}
\caption{Kundagrami \it{et al.}, JCP}
\label{cartoon}
\end{figure}

\begin{figure}[p] \centering
\hspace*{0cm}{\epsfxsize= 16cm \epsfbox{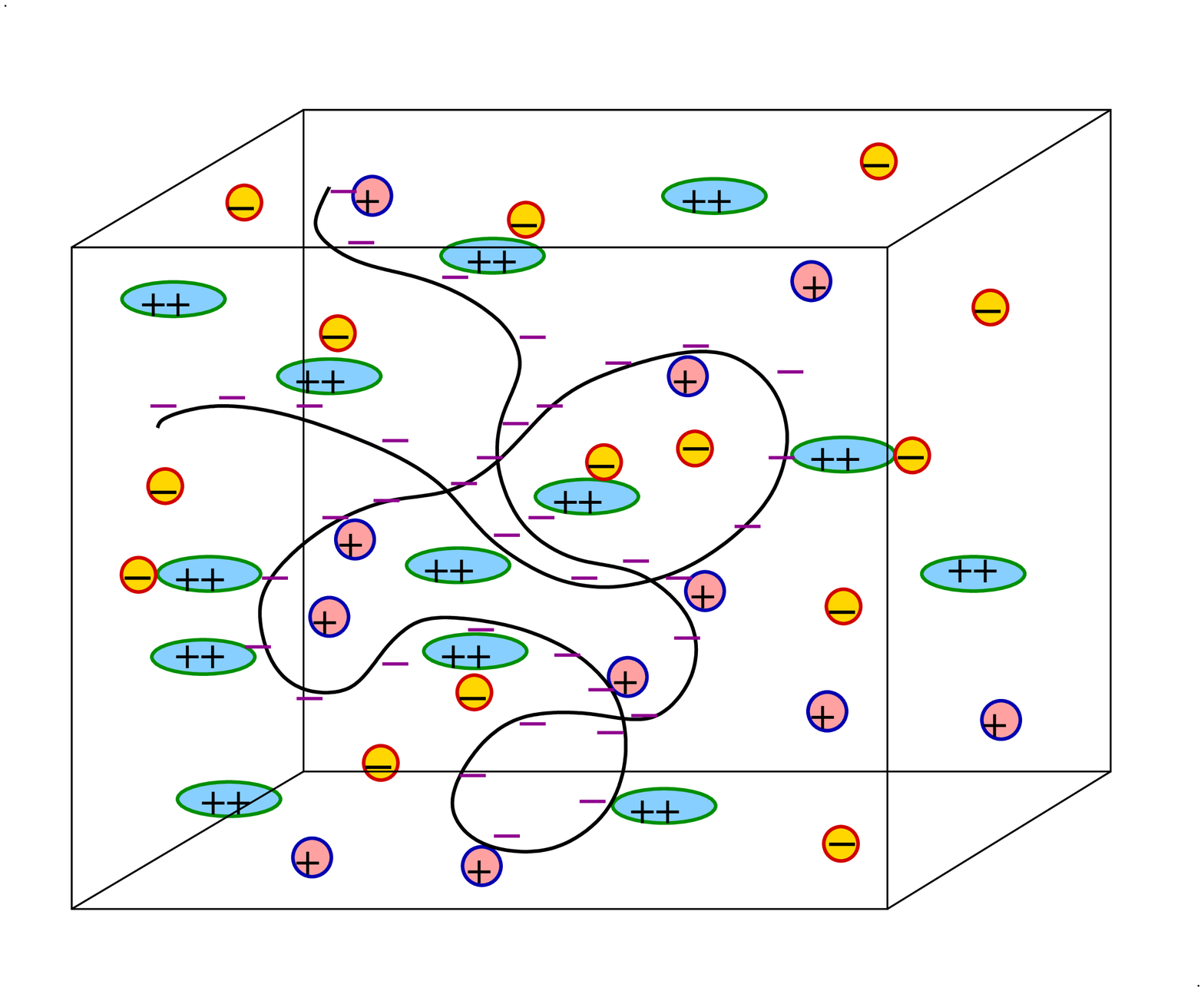}}
\bigskip
\vspace*{0cm}
\caption{Kundagrami \it{et al.}, JCP}
\label{schematic}
\end{figure}

We have proposed earlier a tentative state diagram (Fig. \ref{cartoon}) of the
total charge $f$ ( or the degree of ionization 
$= 1 - \alpha_1 - 2 \alpha_2 + \alpha_3)$ of the 
polyelectrolyte. In this subsection, we present
the actual state diagrams calculated from our theory, as functions of three
major variables - the Bjerrum length ($l_B$), divalent salt concentration 
($c_{s2}$) and the dielectric mismatch parameter ($\delta$). 
In what follows, one of these
variables is fixed and the diagram of states (regions of negative and positive
degree of ionization) is calculated numerically as functions of the other
two. Figs. \ref{lb-delta-cs2-phase}-\ref{cs2phase} describe the complete limiting 
charged states, parts of which have already been discussed in detail in 
preceding subsections.

In Fig. \ref{lb-delta-cs2-phase}(a), the calculated state diagram at  
$\tilde{l_B} = 3.0$ is presented as a function of the divalent salt 
concentration $\tilde{c_{s2}}$ and the dielectric mismatch parameter $\delta$. The
state diagram is qualitatively similar to the proposed one [Fig. \ref{cartoon}],
with the strength of the Coulomb interaction being represented by $\delta$
[Eq. (\ref{f4final})], and with $\tilde{l_B}$ being fixed. To explain the 
diagram, we first choose a specific value of 
$\delta = 2.5$ (see Figs. \ref{delta-1p5-2p5-alphas-l1t}(b) and 
\ref{delta-2p5-cs1-cs2}) and monitor the charged state with increasing divalent 
salt concentration. For low salt, there are not enough divalent 
counterions (Ba$^{++}$) to neutralize the chain and the polyelectrolyte
preserves its sign of charge (state A) of salt-free conditions. At around
$c_{s2} \sim \rho/2$, which is half the monomer concentration, the charge
of the polymer becomes zero (on the locus of first isoelectric points 
- the solid line).
If $c_{s2}$ is increased further, the polymer charge is reversed (state B),
and at around $c_{s2} \sim \rho$, almost all monomers are neutralized by divalent
counterions. The charge reversal is maximum at around this point (on the locus of 
maximum overcharging points
- the dotted line). With $c_{s2}$ increasing even further, more negative 
coions (Cl$^-$) are available in the solution and some of them condense on the 
monomer-Ba$^{++}$ ion-pairs to reduce the degree of overcharging (state C).
The first isoelectric points between states A and B are reached at a higher 
$c_{s2}$
for a lower $\delta$, because a higher fraction of divalent counterions would
remain dissociated in the solution due to a lower Coulomb energy gain. For values of 
$\delta$ higher than  $\simeq 3$, a substantial fraction of monovalent 
counterions of the polymer (Na$^+$) too remain condensed on the chain and the chain 
charge is neutralized with fewer divalent counterions. If $\delta$ is less than  
$\simeq 1.7$, the state 
of overcharging (state B) is never reached and with increasing salt 
concentration, the polymer charge goes through a minimum (on the locus of 
points of minimum charge - the dashed line) before increasing again due to
Cl$^-$ ion condensation. The line of minimum charge (for $\delta$ less than  
$\simeq 1.7$) 
expectedly continues to be the line of maximum charge reversal (for 
$\delta$ greater than $\simeq 1.7$). For very
high salt concentrations, the Coulomb interaction is progressively screened
and all condensed ions begin to rejoin the solution (not included in the state 
diagram). We must, however, be cautioned
that the Debye-H\"{u}ckel (DH) approximation (and consequently $F_3$ in 
Eq. (\ref{f3final})) might not be valid at this high salt regime. A salt
concentration for which the Debye length is equal to the Bjerrum length
($\kappa^{-1} \ge l_B$) can be tentatively set as the highest limit of validity
of the DH theory. For divalent salts it turns out to be 
[see Eq. (\ref{kappasquare2})]
\bea
\label{dhvalid}
c_{s2-max} \simeq \left( 24 \pi l_B^3\right)^{-1}.
\eea 
The steepness of the state boundary (the locus of second isoelectric points - 
the dot-dashed line) implies that the polymer charge becomes zero again
(only applicable for $\delta > 1.7$) due to re-dissolution of condensed ions 
at least an order higher salt concentrations.    
 
The state diagram as a function of $l_B$ and $c_{s2}$ for a fixed value
of $\delta=2.5$ is presented in Fig. \ref{lb-delta-cs2-phase}(b). The diagram is 
qualitatively similar to the previous one, although a much higher salt 
concentration (note the difference in the scale of the coordinate) is
needed to reach the line of minimum charge (dashed) and the line of maximum
overcharging (dotted) at low values of $\tilde{l_B}$ (higher temperatures). 
In this regime, electrostatics becomes progressively weaker with 
increasing temperature, and consequently lower fractions of available divalent 
ions condense.
Regarding this diagram too, the degree of ionization and overcharging (absolute value 
of $f$) can be obtained for the particular value of $\tilde{l_B} = 3.0$ from
Fig. \ref{delta-1p5-2p5-alphas-l1t}(b). 

The state diagram as a function of $\tilde{l_B}$ and $\delta$ for a fixed salt
concentration $\tilde{c_{s2}} = 0.0005$ (equal to the monomer concentration) is
presented in Fig. \ref{cs2phase}.  One essential characteristic is that
the degree of ionization remains steadily at zero (state C1) above a certain 
value of 
$\tilde{l_B}$ (i.e., below a certain temperature), because the Coulomb attraction
is strong enough to make form the monomer-Ba$^{++}$-Cl$^-$ ion-triplet on every
monomer location. This critical value of $\tilde{l_B}$ (on the lower boundary of 
the zero charge state - the dot-dot-dashed line) decreases with higher values of 
$\delta$ 
(higher electrostatic energy gain). The magnitude of the degree of ionization 
as a function of $\tilde{l_B}$ for fixed values of $\delta$ can be obtained
in Fig. \ref{cs2-delta-lb}, and as a function of $\delta$ for a fixed value
of $\tilde{l_B}$ in Fig. \ref{cs2-delta}. Both figures can be analyzed in
conjunction with this state diagram. One small point worthy of note is that 
state D, although implying a non-zero degree of ionization of the same 
sign of the bare polymer charge, has virtually negligible charge in it that 
can be assumed to be zero.  

\subsection*{G. The bridging scenario : a simple theory}

In all previous discussion, we did not consider the bridging configuration 
of nonbonded monovalent monomers
by divalent counterions. This ion-bridging phenomenon is electrostatic in 
nature and can significantly affect polyelectrolyte conformation if present.
In this subsection we will give preliminary results of a simple theory based on 
our model.  We assume that a fraction of condensed divalent ions participate in
bridging. Therefore, when bridging is included the minimization of the free 
energy, as it will turn out, is with respect to five variables. Now, the bridge 
formed by divalent counterion tantamounts to a cross-link
junction of functionality four\cite{liuetal-03}, that in turn can be treated as
an attractive two-body interaction of local nature (like two-body excluded volume 
interaction). Therefore, in the presence of bridging effects due
to divalent counterions, $w$ in Eq. (\ref{f5final}) is replaced by,
\bea
\label{wbridge}
w' = w+\frac{E_{br}}{k_B T} \alpha_{2b},
\eea
where $\alpha_{2b}$ is the ratio of the number of divalent ions that participate 
in bridging to the number of monomers (i. e., $\alpha_{2b} = M_{2b}/N$, 
where $M_{2b}$ is the total number of divalent 
ions involved in bridging). $E_{br}$ is the attractive energy associated with
one bridge, and hence is negative. To calculate $E_{br}$, the relevant dielectric 
constant should be the local one ($\epsilon_l$) since the divalent cation in the
monomer-cation-monomer charge complex sits
between and in the vicinity of both monomers. With the definition\cite{mut-04}
 $\delta=\epsilon l/\epsilon d$, where the distance between both ion pairs in 
the complex remains to be $d=l$ (same as other condensed pairs), it turns out 
that
\bea
\label{ebridge}
E_{br} &=& -\frac{4 e^2}{4 \pi \epsilon_l d} 
+ -\frac{e^2}{4 \pi \epsilon_l . 2 d} \nonumber  \\
&=& - \frac{7}{2} \tilde{l_B} \delta k_B T.
\eea
We must add the third virial term in the chain free energy ($F_5$, 
Eq. (\ref{f5final})) to maintain stability in the system in case of a negative
$w'$. Combining Eqs. (\ref{wbridge}) and (\ref{ebridge}), therefore, the chain
free energy takes the form,
\bea
\label{f5finalbridge}
\frac{F_5}{N k_B T} &=& \frac{3}{2 N}\left( \tilde{l_1} - 1
- \log \tilde{l_1} \right) + \frac{4}{3}\left( \frac{3}{2 \pi} \right)^{3/2}
(w - \frac{7}{2} \tilde{l_B} \delta \alpha_{2b}) \frac{1}{\sqrt{N}} 
\frac{1}{\tilde{l_1}^{3/2}} \\ \nonumber
&+& \frac{1}{N} \frac{w_3}{\tilde{l_1}^3} + 2 \sqrt{\frac{6}{\pi}} f^2
\tilde{l_B} \frac{N^{1/2}}{\tilde{l_1}^{1/2}} \Theta_0 (a),
\eea  
where $w_3$ is the third-virial coefficient which is necessarily positive.
Further we note that if a fraction $\alpha_{2b}$ of condensed divalent counterions 
participate in bridging, a 
fraction $\alpha_{2a}=\alpha_2 - \alpha_{2b}$ does not. Therefore, the 
electrostatic energy related to the formation of monomer-cation monocomplexes
($F_4$, Eq. (\ref{f4final})) is modified after the inclusion of bridging 
interaction as,
\bea
\label{f4finalbridge}
\frac{F_4}{N k_B T} = - \alpha_1 \delta \tilde{l_B} 
- 2 (\alpha_{2a} - \alpha_3) \delta \tilde{l_B} - \alpha_3 \delta_2 \tilde{l_B},
\eea 
where $\alpha_3 \le \alpha_{2a}$.
The other parts of the free energy remain unaltered and are given as: $F_1$ in
Eq. (\ref{f1final}), $F_2$ in Eq. (\ref{f2final}), and $F_3$ in 
Eq. (\ref{f3final}). In all these cases, $\alpha_2 = \alpha_{2a} + \alpha_{2b}$.

The total free energy $F=F_1 + .. + F_5$ is minimized now for a new set of 
five variables, 
$\alpha_1, \alpha_{2a}, \alpha_{2b}, \alpha_3$ and $\tilde{l_1}$, and the 
polymer and counterions are free to explore every possible degree of freedom. 
The representative result
is given in Fig. \ref{lb-delta-bridge-cs2}. The parameters chosen are:
$N=100, \tilde{\rho} = 0.0008, \tilde{l_B} = 3.0, \delta=1.9, \tilde{c_{s1}} = 0, 
w=2.0, w_3=0.25$. For very low divalent salt concentrations, the 
conformations are very similar to the case where bridging is absent. At modest
temperatures ($\tilde{l_B} = 3.0$ in water) and for low salt ($\tilde{c_{s2}} 
< 0.00027$), almost all added divalent counterions condense, but they from
monocomplexes (no bridging, Fig. \ref{lb-delta-bridge-cs2}(a)). At a particular
$\tilde{c_{s2}}^*$, which depends on the prevalent physical conditions, all
divalent ions suddenly form dicomplexes (bridging) accompanied by a collapse 
of the chain [Fig. \ref{lb-delta-bridge-cs2}(b), in which $\tilde{l_1} \ll 1$ 
for $\tilde{c_{s2}}> 0.00027$] and a huge gain in electrostatic bridging free 
energy [Fig. \ref{lb-delta-bridge-cs2}(c)]. The Cl$^-$ ions condense as they do 
for no-bridging scenario only if the divalent salt concentration is lower than 
the collapse concentration $\tilde{c_{s2}}^*$. Above that, Cl$^-$ ions become 
free as every condensed divalent cation is attached to two monomers. The effect 
of the excluded volume parameter $w$ is evident in Eqs. (\ref{wbridge}) and 
(\ref{f5finalbridge}) as they show that a higher $w$ will require a higher Coulomb
strength or divalent salt concentration to effect the bridging collapse. Until
the collapse, the distribution of counterions and the polymer conformations are
quite similar to that of the  'no-bridging' cases 
(see Fig. \ref{delta-1p5-2p5-alphas-l1t} for example). This explains our 
choice of $w=0$ for the rest of the article (except for this subsection). In the 
'no-bridging' scenario, different positive values of $w$ would only render minor 
qualitative changes to our results. 

We further notice that the 
collapse concentration, $\tilde{c_{s2}}^*$, decreases with increasing Coulomb
strength [Fig. \ref{lb-delta-cs2trans}(a),(b)] confirming that the first order 
collapse induced due to ion-bridging by divalent (or multivalent) cations is an 
electrostatic phenomenon. In addition, Fig. \ref{lb-delta-cs2trans}(c) shows 
that $\tilde{c_{s2}}^*$ roughly varies inversely with both forms of Coulomb 
strength, $\tilde{l_B}$ and $\delta$.
This is a remarkable prediction for experimentalists and we find 
$\tilde{l_B} \tilde{c_{s2}}^* \simeq 0.0006$ (for $\delta=2.5$) and 
$\delta \tilde{c_{s2}}^* \simeq 0.0005$ (for $\tilde{l_B}=3.0$).   

In conclusion, our model predicts a bridging
transition, which we believe depends sensitively on temperature and dielectric
heterogeneity, as well as on the availability of divalent counterions. At this point,
we leave the rest of the bridging analysis for future publication.

\section{Conclusions}

\setcounter{section}{4}
\setcounter{equation}{0}

We have extended Muthukumar's adsorption theory\cite{mut-04} for condensation of 
monovalent counterions on a flexible polyelectrolyte by including  
divalent counterions. It is observed that the divalent counterions 
replace the monovalent ones in the competitive adsorption
process. For moderate values of the dielectric mismatch parameter ($\delta$),
temperature and bulk dielectric constant ($l_B$), numbers of both condensed 
divalent cations (Ba$^{++}$)
and monovalent anions (Cl$^-$) increase and the number of condensed monovalent 
cations (Na$^+$) decrease monotonically with increasing concentration of added
divalent salt (BaCl$_2$). As observed in previous theories, experiments, and 
simulations, a moderate amount of divalent salt entirely compensates the
polyelectrolyte (NaPSS) charge and consequently contracts the chain to its
Gaussian size.  The divalent salt concentration at which the charge neutralization
takes place is roughly half the salt concentration implying that almost
all divalent cations added to the solution condense at those 
modest parameter values. With further increase of divalent salt
concentration, condensed divalent counterions overcharge the chain 
resulting in reswelling of the chain. This phenomenon is addressed
for the first time theoretically for a flexible chain allowed to take all 
possible conformations. The charge reversal, however,
is absent at modest temperatures regardless of the salt concentration if the 
dielectric heterogeneity is weaker than a critical value. The minimum degree
of heterogeneity ($\delta$) required to condense all available divalent 
cations at a particular temperature and solution ($l_B$) increases with 
temperature.
At room temperature in aqueous solutions ($l_B \sim 7\AA$), the adsorption
theory predicts that the ratio of the bulk to local dielectric constant must 
be at least 1.7 to overcharge the polyelectrolyte.

It can be conceived that there are three length scales involved in the 
polyelectrolyte system analyzed by this adsorption theory. They are, the Bjerrum 
length $l_B$ (representing the equilibrium temperature $T$ of a solution with
bulk dielectric constant $\epsilon$), the
Debye length $\kappa^{-1}$ (representing the salt concentration $c_{si}$),
and the third one related to the strength of the dielectric heterogeneity in
the vicinity of the polyelectrolyte backbone ($\delta$). We have developed 
the diagrams of charged states of the polymer in terms of these three variables.  
Both higher $\delta$ (lower ratio of local to bulk $\epsilon$) and $l_B$ 
(lower temperature or bulk $\epsilon$) facilitate ion 
condensation of all types. Consequently, the isoelectric point (charge 
neutralization) and subsequent charge inversion are achieved at 
lower temperatures for higher $\delta$ and vice versa. The absolute value of 
the maximum charge reversal and simultaneous maximal reswelling increase with 
$\delta$.  For a minimal dielectric mismatch ($\delta \sim 1$), no overcharging 
is predicted for any temperature. This limit is the closest comparable to
the simulations and it explains why in some simulations overcharging
is never observed\cite{liuetal-03}. 

For moderate values of $\delta$ ($\ge 2$), maximal overcharging and reswelling
occurs at intermediate temperatures lower than that of the first 
isoelectric point (at which a moderate amount of divalent cations and a fraction
of monovalent anions neutralize the chain). If temperature is further decreased
($l_B$ increased), progressively higher fraction of monovalent anions (Cl$^-$)
condense reducing the degree of overcharging. At low enough temperatures,
regardless of the degree of dielectric heterogeneity, the second and trivial 
isoelectric point (at which all  monomers are neutralized by the presence of
both Ba$^{++}$ and Cl$^-$ ions) is achieved and the Gaussian statistics is 
reestablished. 

We have determined the physical limits of the strength parameter for the formation
of monomer-Ba$^{++}$-Cl$^-$ triplets as functions of the basic dielectric mismatch
strength $\delta$. It has been shown that the qualitative picture of the total 
effective  
charge and conformation of the polyelectrolyte remain unaltered although the 
distribution of counterions and coions around it may vary for a change in the 
triplet strength 
parameter ($\delta_2$) within these limits. For an unrealistic value of the 
parameter closer to 
its higher limit, increased Cl$^-$ condensation may marginally eliminate 
overcharging at all physical conditions.  

Analysis of the free energy hightlights the competition between the entropic 
contribution of dissociated ions and the electrostatic contribution of condensed
ions in shaping the counterion distribution around the polyelectrolyte. The 
local dielectric constant and temperature have similar contribution to the
electrostatics of the system (except at very high salt concentrations). If 
electrostatic bridging of non-bonded monomers 
by divalent counterions is effective due to a weak excluded volume effect,
there is a first order collapse at modest conditions. The transition salt 
concentration varies approximately inversely with the Coulomb strength at modest
conditions. In summary,
our theory predicts that both electrostatics and the dielectric inhomogeneity near 
the backbone of the polymer (reflecting the nonuniversal chemical nature of the
system) are responsible for charge reversal in polyelectrolyte systems.      
 
\section{Acknowledgements}

Financial support for this work was provided by the NIH Grant No. 1R01HG002776-01,
National Science Foundation (NSF) Grant No. 0605833, and MRSEC at the University 
of the Massachusetts, Amherst.  Fruitful discussions with Zhaoyang Ou are 
gratefully acknowledged.

%
%

\section*{FIGURE CAPTIONS}
\begin{figure}[p] \centering
\hspace*{0cm}{\epsfxsize= 16cm \epsfbox{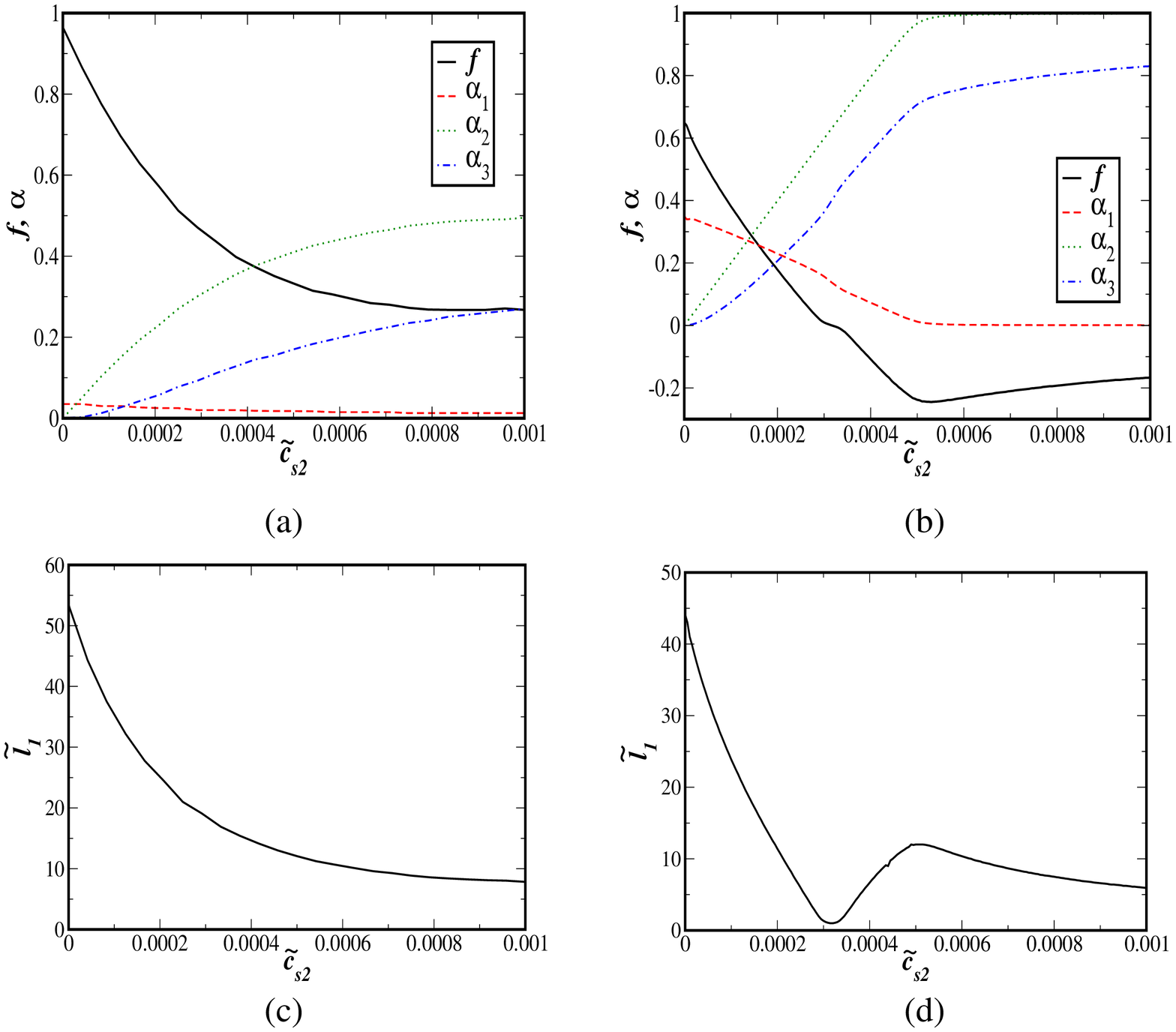}}
\bigskip
\vspace*{0cm}
\caption{Kundagrami \it{et al.}, JCP}
\label{delta-1p5-2p5-alphas-l1t}
\end{figure}

\pagestyle{empty}
\begin{description}

\item[Fig. 1] Sketch of charged states for an isolated polyelectrolyte chain 
(of NaPSS type) in dilute solutions (water) in the presence of a divalent salt 
(of BaCl$_2$ type) 
as functions of the Coulomb strength ($l_B \sim (\epsilon T)^{-1}$) and the salt 
concentration. Points left to the isoelectric line (on which net effective
charge (degree of ionization) on the polymer is zero) correspond to states
in which the sign of polymer charge is unchanged (negative). However, there
is a locus of points for intermediate values of salt concentration at which 
the net charge is a minimum.   
Right to the isoelectric line the effective polymer charge is reversed (positive). 
If the isoelectric point is crossed along the line of minimum
charge from left to right, it becomes the line of maximum overcharging. Dashed
part of the isoelectric line is beyond the Debye-H\"{u}ckel limit.

\item[Fig. 2] Schematic diagram of the system consisting of the isolated 
polymer chain, condensed counterions, dissociated mobile ions, and the 
solution as the background interacting only through the dielectric constant
$\epsilon$. Possible charge complexes for each monomer: monomer (-1), 
monomer-monovalent (-1,+1), monomer-divalent (-1,+2), and 
monomer-divalent-monocoion (-1,+2,-1). The dielectric constant $\epsilon_l$ 
in the vicinity of the chain 
is much lower than the bulk value. To reach equilibrium the major 
competition is between the translational entropy of the dissociated ions
and the Coulomb energy gain of the condensed ions.    

\item[Fig. 3] Competitive displacement of monovalent counterions by divalent 
counterions: fraction of condensed ions ($\alpha_1, \alpha_2, \alpha_3$), 
degree of ionization ($f = 1 - \alpha_1 - 2 \alpha_2 + \alpha_3$) in (a) and (b)
for $\delta=1.5$ and 2.5 respectively, and
the expansion factor ($\tilde{l_1} = 6 R_g^2/N l^2$) in (c) and (d) for the same
$\delta$ values plotted against
the divalent salt concentration($\tilde{c_{s2}} \sim 0.01 c_s (\mbox{M}))$. 
For Gaussian chain, $\tilde{l_1} = 1$. Other parameters are: $N = 1000,
\tilde{\rho} = 0.0005, \tilde{l_B} = 3.0, w = 0$, and $\tilde{c_{s1}} = 0$. 
Note, for lower
$\delta$, there is no overcharging. For higher $\delta$, almost 
all divalent counterions condense on the chain replacing the monovalent ones. 
The number of condensed negative coions ($\alpha_3$) closely follow $\alpha_2$
for this particular value of $\delta$. The sign of $f$ is reversed 
({\it overcharging}) at some
concentration of BaCl$_2$. Near the isoelectric point ($f \sim 0$), the chain
is Gaussian due to minimal electrostatic repulsion. It swells due to
overcharging if $\tilde{c_{s2}}$ is further increased.

\item[Fig. 4] Effect of valency of counterions: comparison of degree of 
ionization ($f$) in (a), and size expansion factor
($\tilde{l_1}$) in (b), of the polyelectrolyte in presence of either monovalent 
or divalent salt. $\delta=2.5$ and other parameters are the same as in Fig. 
\ref{delta-1p5-2p5-alphas-l1t}.
$\tilde{c_{s2}} = 0$ is zero when $\tilde{c_{s1}}$ is added and vice versa.
Divalent counterions can neutralize and consequently collapse the polymer at
moderate conditions of $\tilde{l_B} = 3.0$. For monovalent counterions, collapse 
(to Gaussian chain)
is only possible at very low temperatures. If $\tilde{c_{s2}}$
is increased beyond the isoelectric point, the chain expands due to
overcharging. 

\item[Fig. 5] Dependency of overcharging on $\tilde{l_B}$: degree of ionization 
($f$) in (a), and size expansion factor ($\tilde{l_1}$) in (b), of the 
polyelectrolyte plotted against $\tilde{l_B}$ for different values of $\delta$.
Parameters are: $N=1000, \tilde{\rho} = 0.0005, \tilde{c_{s1}} = 0$, and
$\tilde{c_{s2}} = 0.0005$. Collapse and subsequent overcharging occur for higher
values of $\delta$. Isoelectric point is reached at lower $l_B$
for higher values of  $\delta$. This overcharging behavior contrasts with 
Figs. 2(a) and 3 in Ref.\cite{mut-04} for monovalent salts. 

\item[Fig. 6] Dependency of overcharging on $\delta$: fraction of condensed 
ions ($\alpha_1, \alpha_2, \alpha_3$),
degree of ionization ($f$) in (a), and
the expansion factor ($\tilde{l_1}$) in (b), plotted against
the $\delta$-parameter for $\tilde{l_B} = 3.0$. 
Other parameters are the same as in Fig. \ref{cs2-delta-lb}.
Charge neutralization occurs for $\delta$ about 1.7; overcharging is possible
only if $\delta > 1.7$. Cl$^-$ condensation ($\alpha_3$) is higher for 
higher values of $\delta$, and that reduces overcharging.

\item[Fig. 7] Effect of dielectric mismatch for the coion: fraction of condensed 
Cl$^-$ ions ($\alpha_3$), 
degree of ionization $f$ in (a) and (b) for $\delta=2.5$ and 1.5 respectively, 
and expansion factor  ($\tilde{l_1}$) in (c) and (d) for the same $\delta$ 
values plotted against $\delta_2$ for its whole physical range possible 
for the respective value of  
$\delta$.  Other parameter values are the same as in Fig. \ref{cs2-delta}.
For higher $\delta=2.5$, progressively higher values of $\delta_2$ 
reduce and finally eliminate overcharging (and consequently the reexpansion 
of the chain) by increasing $\alpha_3$. 
For $\delta = 1.5$ (no overcharging at any temperature), 
fraction of both condensed divalent counterions and monovalent coions, $\alpha_2$ 
and $\alpha_3$ respectively, increase to leave $f$ (and $R_g$) 
approximately unchanged.  

\item[Fig. 8] Contributions to free energy: separate parts of the free energy 
for a fixed divalent salt
concentration ($\tilde{c_{s2}} = 0.0005$) as functions of (a) Bjerrum 
length $\tilde{l_B}$ and (b) dielectric mismatch strength ($\delta$). For
(a), $\delta = 2.5$ and (b), $\tilde{l_B}=3.0$. Parameters are: $N=1000, 
\tilde{\rho} = 0.0005, \tilde{c_{s1}} = 0$. Energies are: $F_1$ (dot) =
entropy of mobility along the chain, $F_2$ (dash) = translational entropy
of mobile ions, $F_3$ (dot-dash) = Debye-H\"{u}ckel correlation between mobile
ions, $F_4$ (dot-dash-dash) = Coulomb attraction between condensed ions 
and $F_{tot}$ (solid) = total free energy. For fixed salt concentration, the major
competition is between the translational entropy (increase
with temperature) and Coulomb attraction (increase with
both $\tilde{l_B}$ and $\delta$). The similarity in the roles of $\tilde{l_B}$ 
and $\delta$ is evident. 

\item[Fig. 9] (a) The state diagram of the total charge density of the polymer 
($f$) for $\tilde{l_B} = 3.0$ as functions of the dielectric 
mismatch $\delta$ and the divalent salt concentration $\tilde{c_{s2}}$.
Parameters are: $N=1000, \tilde{\rho} = 0.0005, \tilde{c_{s1}} = 0$. Charged 
states
are: A,D - negative, B,C - positive (note: original polymer charge is
negative). Lines are: isoelectric branch one (solid), maximum overcharging
(dot), isoelectric branch two (dot-dash), and minimum charge (dash);
 (b) the state diagram of $f$ at
a fixed dielectric mismatch strength ($\delta = 2.5$) as functions of the
Bjerrum length $\tilde{l_B}$ and the divalent salt concentration
$\tilde{c_{s2}}$. Other parameters, states, and lines are the same as in (a). 

\item[Fig. 10] The state diagram of the total charge density on the polymer 
($f$) at
a fixed divalent salt concentration ($\tilde{c_{s2}} = 0.0005$) as functions of
the dielectric mismatch strength $\delta$ and the Bjerrum length $\tilde{l_B}$. 
Other parameters are the same as in Fig. 9. Charged 
states 
are: A,D - negative, B,C - positive, and C1 - zero. Lines are:
isoelectric branch one (solid), maximum overcharging (dot), isoelectric branch 
two (dot-dash), minimum charge (dash), and zero charge (dash-dot-dot). 
  
\item[Fig. 11] Effect of bridging: degree of ionization ($f$), fractions of 
condensed ions 
($\alpha$'s) (a), size expansion factor ($\tilde{l_1}$) (b), and
separate parts of the free energy ($F$'s) (c) as functions of divalent salt 
concentration ($c_{s2}$) when monomer-bridging by divalent cations is taken
into account. Parameters are: $N=100, \tilde{\rho} = 0.0008, \tilde{l_B} = 3.0, 
\delta=1.9, \tilde{c_{s1}} = 0, w=2.0, w_3=0.25$. 
Bridging induces a first-order collapse transition with a sudden
gain in electrostatic ion-bridging energy. At the transition salt concentration,
all monomer-divalent cation ion-pairs ('monocomplex'es) give way to 
monomer-cation-monomer ion bridges ('dicomplex'es).  

\item[Fig. 12] Effect of Coulomb strength on collapse: the salt concentration
$\tilde{c_{s2}}^*$, at which the first-order collapse occurs, as functions of 
$\tilde{l_B}$ (a), and $\delta$ (b). In (a), $\delta=2.5$. In (b), 
$\tilde{l_B}=3.0$. All other parameters are the same as in Fig. 11. Lowering
of $\tilde{c_{s2}}^*$ with Coulomb strength indicates that the collapse is due
to electrostatic interactions. $\tilde{l_B}$ and $\delta$ play similar roles, as
expected. In (c): $\tilde{l_B} \tilde{c_{s2}}^* \simeq 0.0006$ 
(for $\delta=2.5$) and
$\delta \tilde{c_{s2}}^* \simeq 0.0005$ (for $\tilde{l_B}=3.0$).

\end{description}

\newpage

\newpage

\newpage
\begin{figure}[ht]  \centering
\begin{minipage}{15cm}
\vspace*{0.0cm}
\hspace*{0.0cm}\textsf{\textbf{(a)}}\\
\vspace*{0.5cm}
\hspace*{0.0cm}{\epsfxsize= 14cm \epsfbox{delta-2p5-f-cs1-cs2.eps}}\\
\vspace*{0.5cm}
\hspace*{0.0cm}\textsf{\textbf{(b)}}\\
\vspace*{0.5cm}
\hspace*{0.0cm}{\epsfxsize= 14cm \epsfbox{delta-2p5-l1t-cs1-cs2.eps}}\\
\caption{Kundagrami \it{et al.}, JCP}
\label{delta-2p5-cs1-cs2}
\end{minipage}
\end{figure}

\newpage
\begin{figure}[ht]  \centering
\begin{minipage}{15cm}
\vspace*{0.0cm}
\hspace*{0.0cm}\textsf{\textbf{(a)}}\\
\vspace*{0.5cm}
\hspace*{0.0cm}{\epsfxsize= 14cm \epsfbox{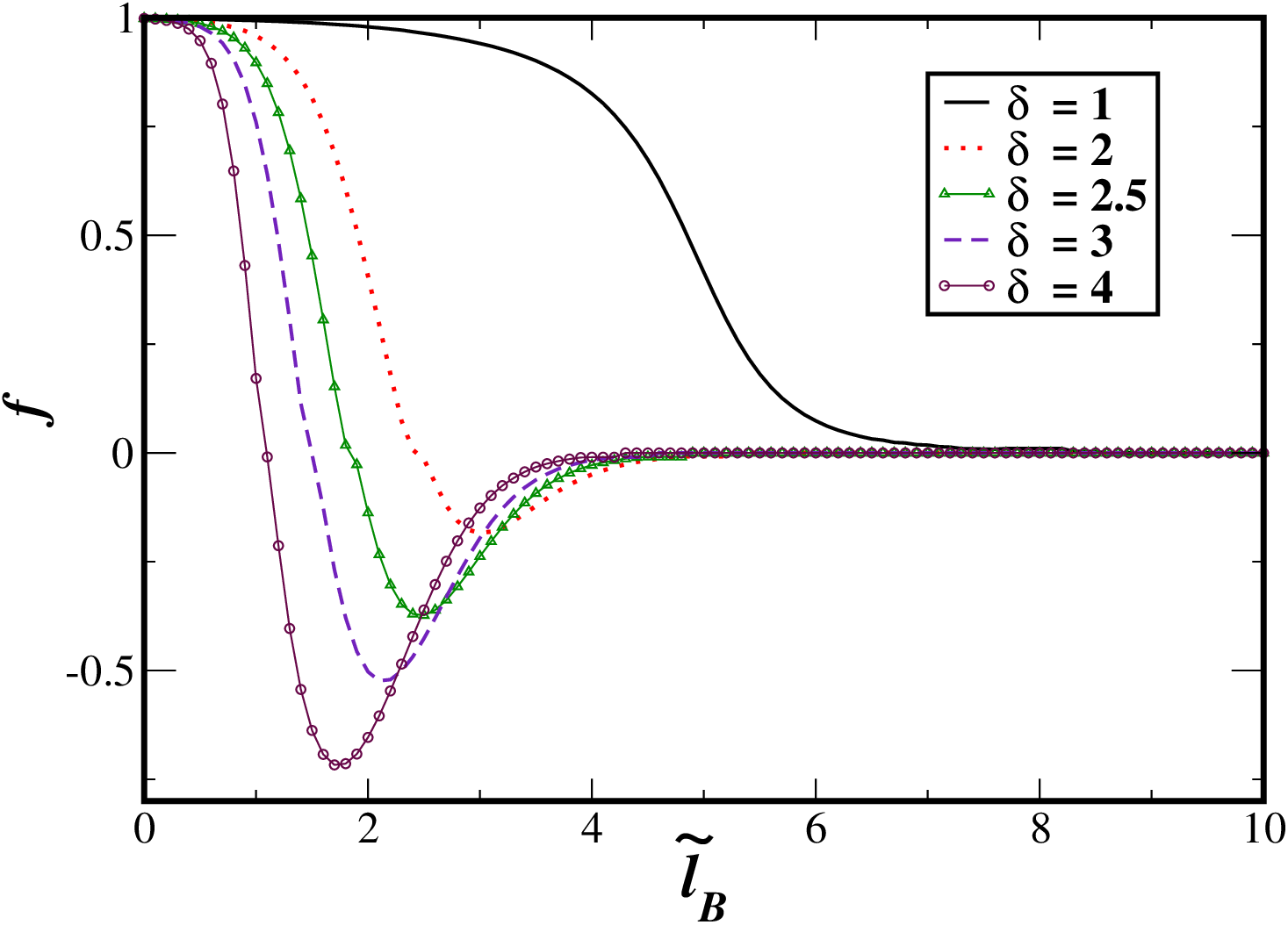}}\\
\vspace*{0.5cm}
\hspace*{0.0cm}\textsf{\textbf{(b)}}\\
\vspace*{0.5cm}
\hspace*{0.0cm}{\epsfxsize= 14cm \epsfbox{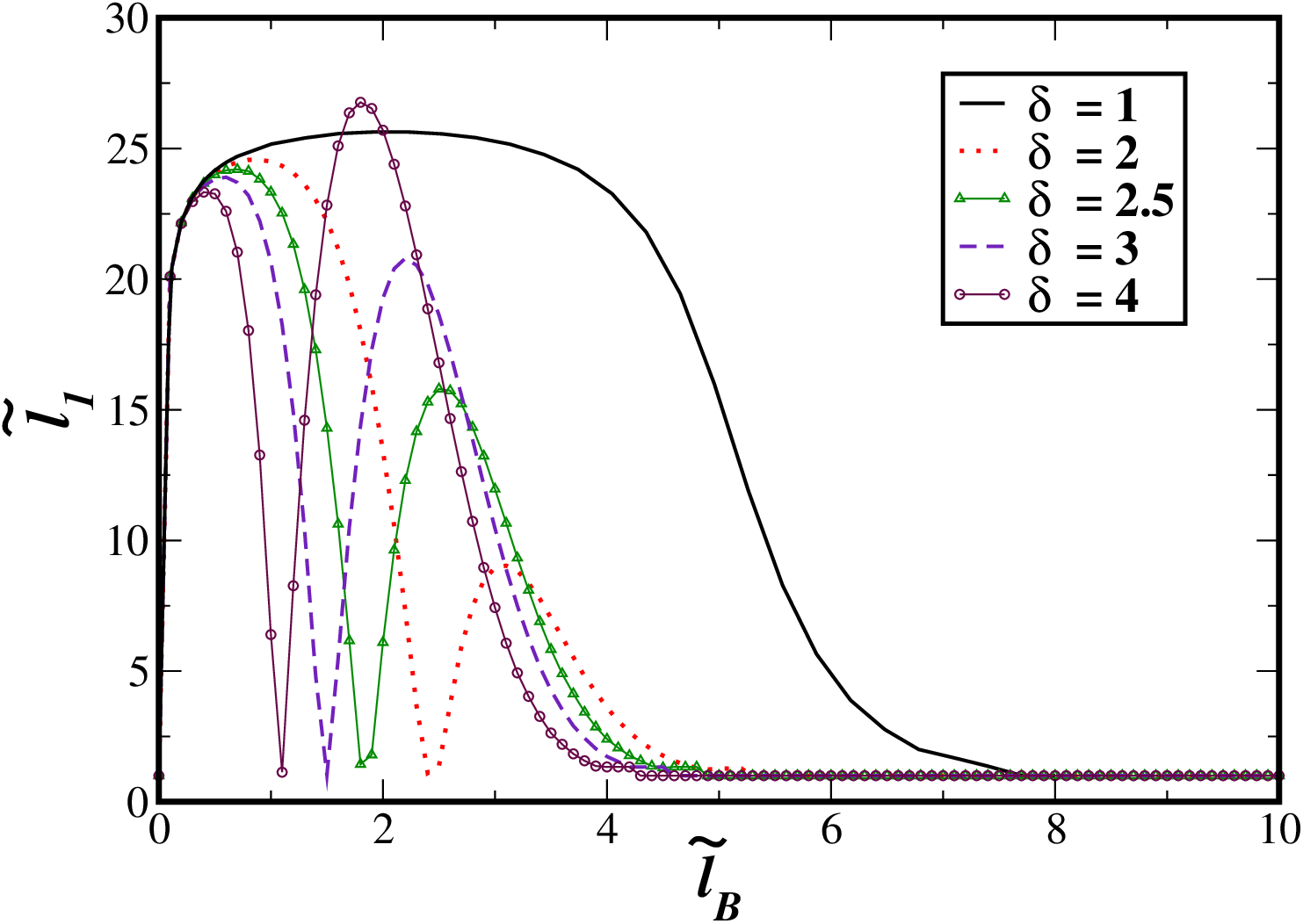}}\\
\caption{Kundagrami \it{et al.}, JCP}
\label{cs2-delta-lb}
\end{minipage}
\end{figure}

\newpage
\begin{figure}[ht]  \centering
\begin{minipage}{15cm}
\vspace*{0.0cm}
\hspace*{0.0cm}\textsf{\textbf{(a)}}\\
\vspace*{0.5cm}
\hspace*{0.0cm}{\epsfxsize= 13cm \epsfbox{cs2-delta-alphas.eps}}\\
\vspace*{0.5cm}
\hspace*{0.0cm}\textsf{\textbf{(b)}}\\
\vspace*{0.5cm}
\hspace*{0.0cm}{\epsfxsize= 13cm \epsfbox{cs2-delta-l1t.eps}}\\
\caption{Kundagrami \it{et al.}, JCP}
\label{cs2-delta}
\end{minipage}
\end{figure}

\begin{figure}[p] \centering
\hspace*{0cm}{\epsfxsize= 16cm \epsfbox{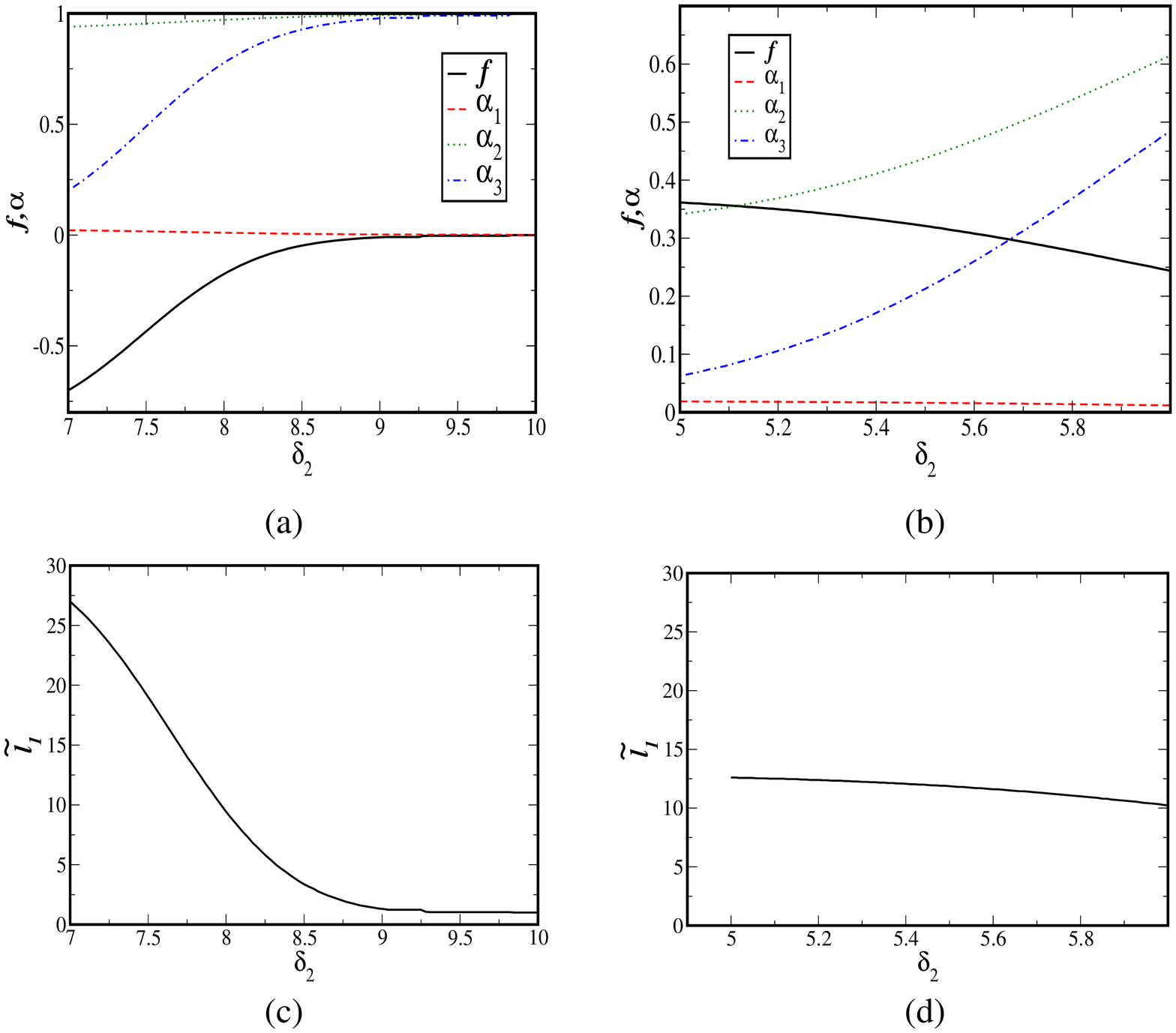}}
\bigskip
\vspace*{0cm}
\caption{Kundagrami \it{et al.}, JCP}
\label{delta2-ovch-noovch-alphas-l1t}
\end{figure}

\newpage
\begin{figure}[ht]  \centering
\begin{minipage}{15cm}
\vspace*{0.0cm}
\hspace*{0.0cm}\textsf{\textbf{(a)}}\\
\vspace*{0.5cm}
\hspace*{0.0cm}{\epsfxsize= 14cm \epsfbox{delta2p5-freen.eps}}\\
\vspace*{0.5cm}
\hspace*{0.0cm}\textsf{\textbf{(b)}}\\
\vspace*{0.5cm}
\hspace*{0.0cm}{\epsfxsize= 14cm \epsfbox{lb3-freen.eps}}\\
\caption{Kundagrami \it{et al.}, JCP}
\label{free-energy}
\end{minipage}
\end{figure}

\newpage
\begin{figure}[ht]  \centering
\begin{minipage}{15cm}
\vspace*{0.0cm}
\hspace*{0.0cm}\textsf{\textbf{(a)}}\\
\vspace*{0.5cm}
\hspace*{0.0cm}{\epsfxsize= 14cm \epsfbox{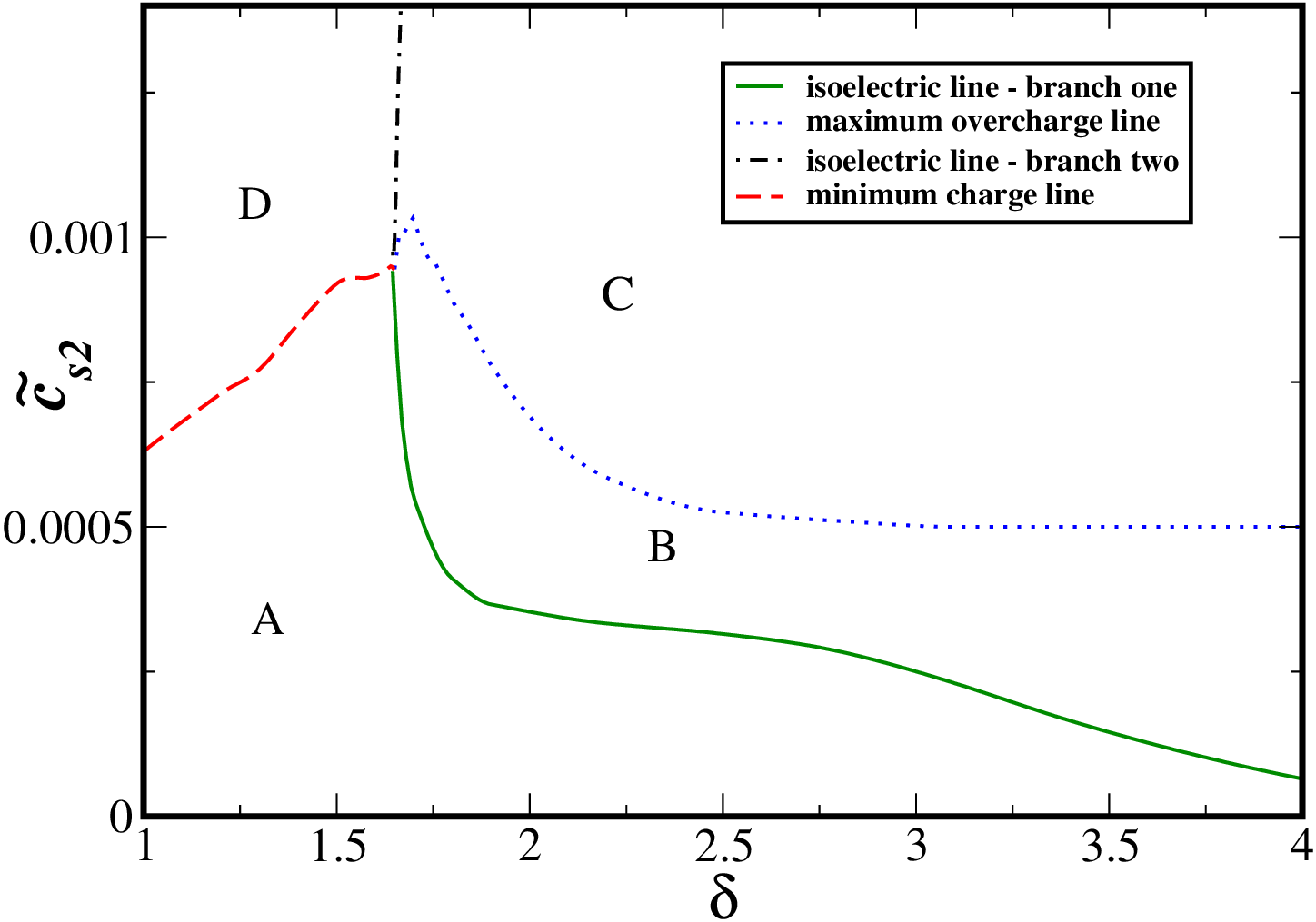}}\\
\vspace*{0.5cm}
\hspace*{0.0cm}\textsf{\textbf{(b)}}\\
\vspace*{0.5cm}
\hspace*{0.0cm}{\epsfxsize= 14cm \epsfbox{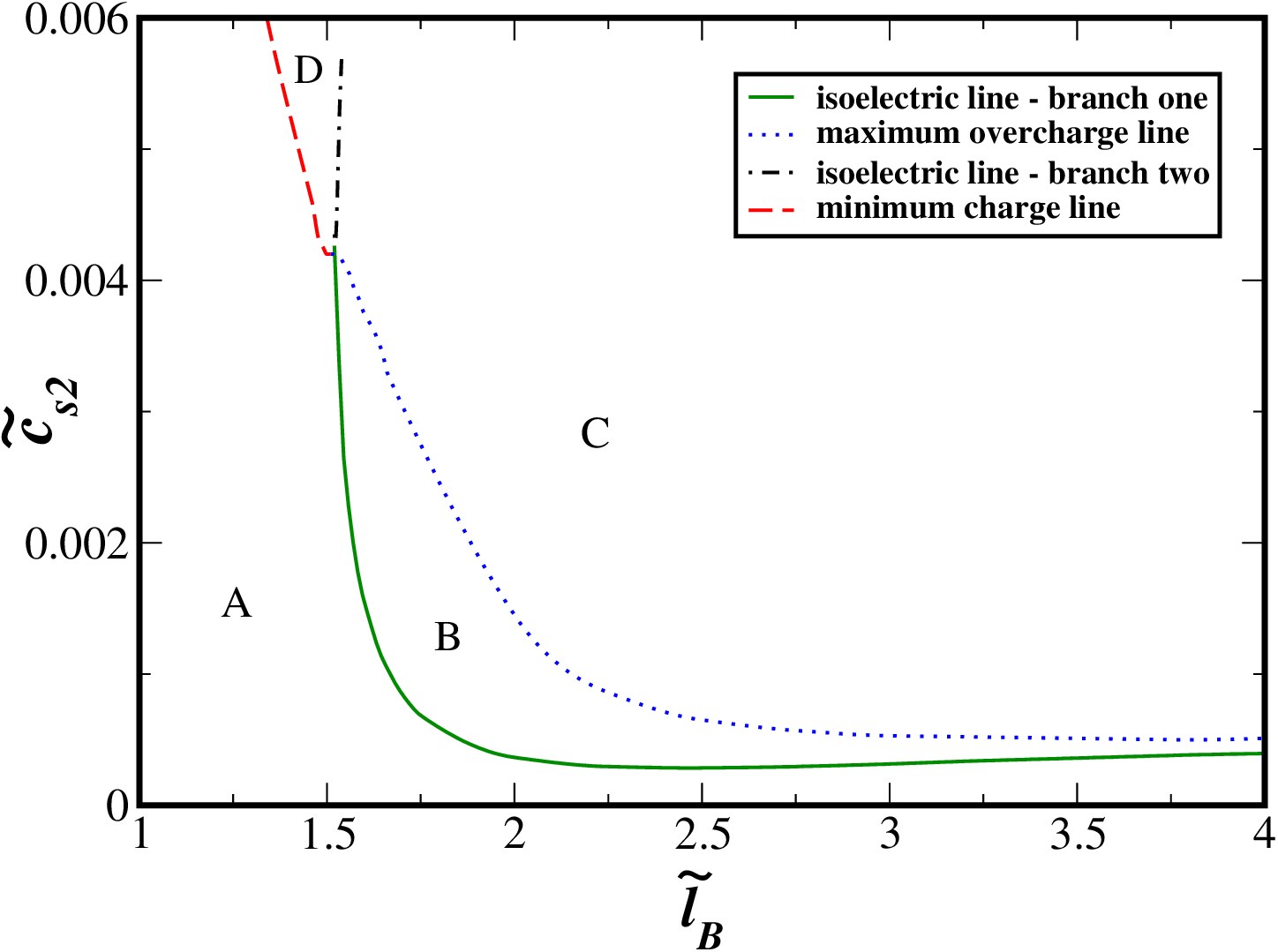}}\\
\caption{Kundagrami \it{et al.}, JCP}
\label{lb-delta-cs2-phase}
\end{minipage}
\end{figure}

\begin{figure}[p] \centering
\hspace*{0cm}{\epsfxsize= 14cm \epsfbox{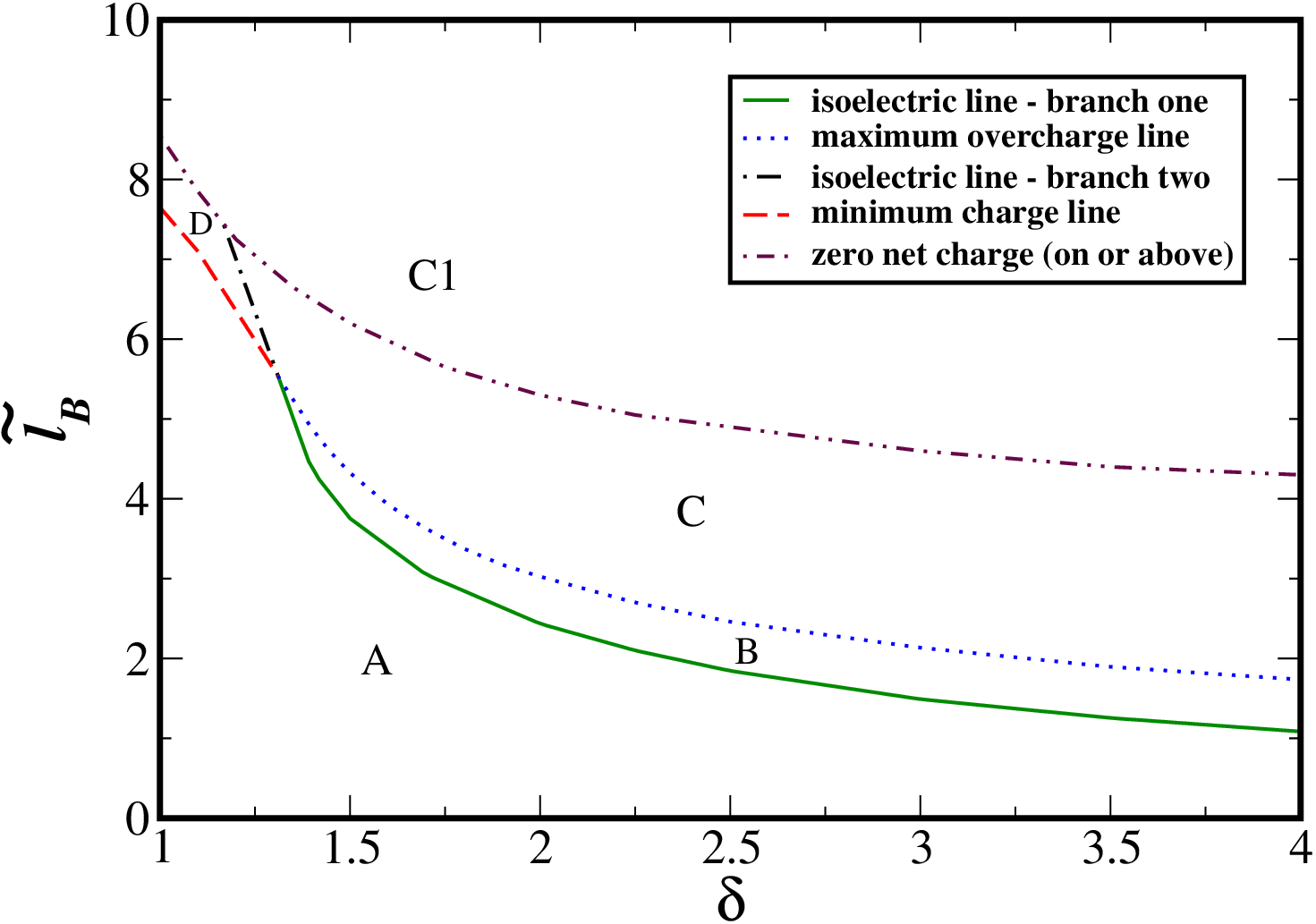}}
\bigskip
\vspace*{0cm}
\caption{Kundagrami \it{et al.}, JCP}
\label{cs2phase}
\end{figure}

\newpage
\begin{figure}[ht]  \centering
\begin{minipage}{16cm}
\vspace*{-2.0cm}
\hspace*{0.0cm}\textsf{\textbf{(a)}}\\
\vspace*{0.5cm}
\hspace*{0.0cm}{\epsfxsize= 10cm \epsfbox{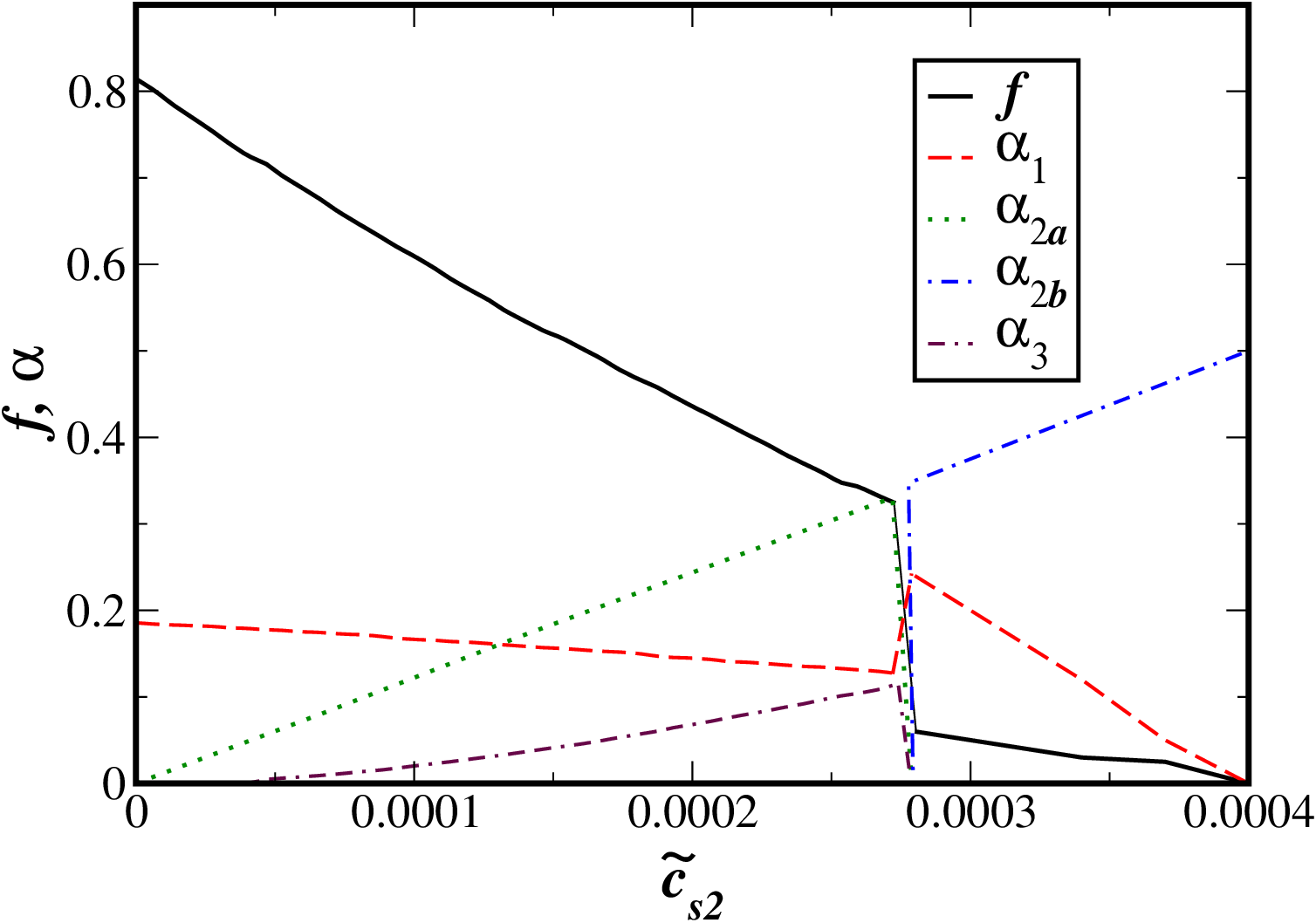}}\\
\vspace*{0.5cm}
\hspace*{0.0cm}\textsf{\textbf{(b)}}\\
\vspace*{0.5cm}
\hspace*{0.0cm}{\epsfxsize= 10cm \epsfbox{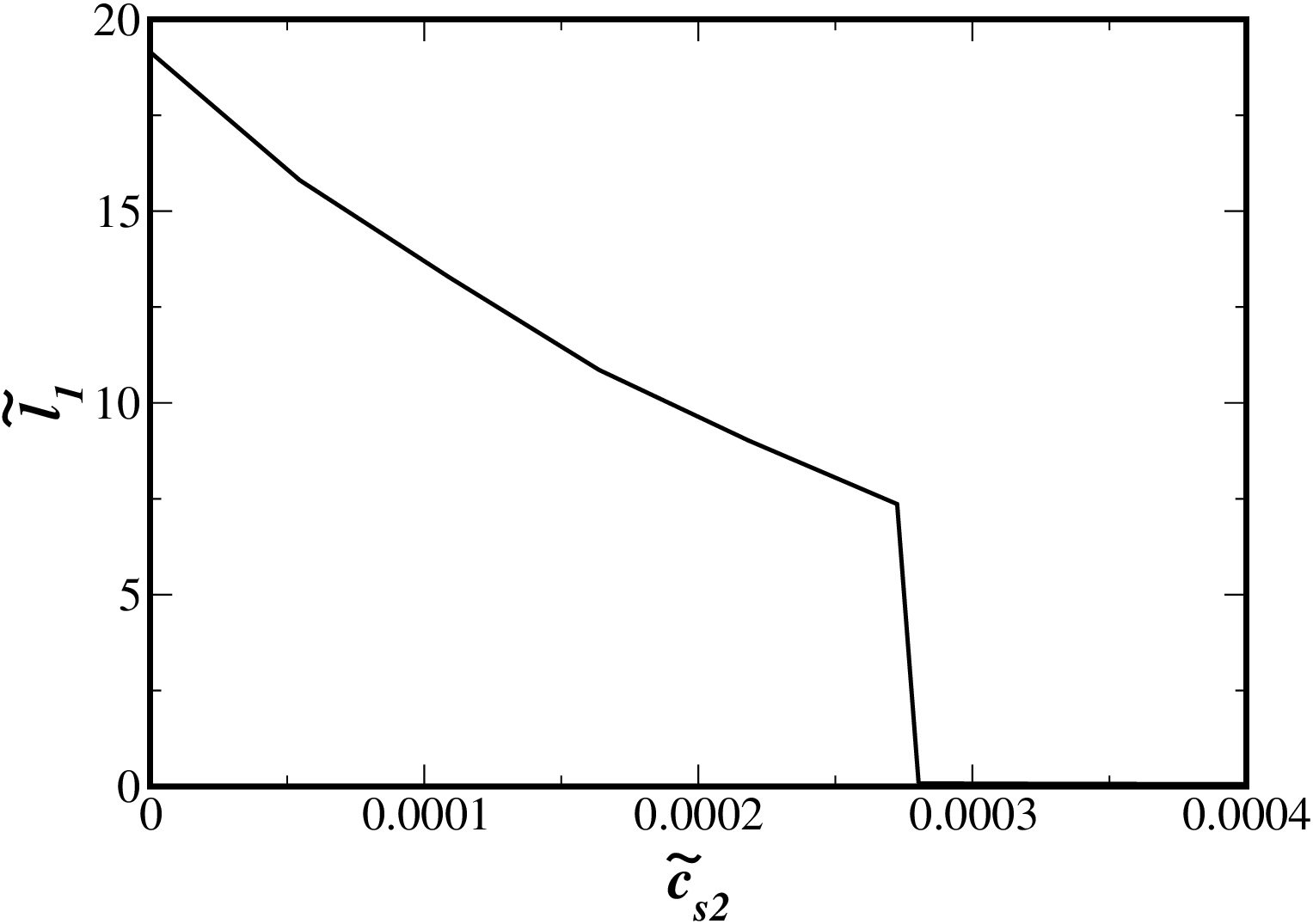}}\\
\vspace*{0.5cm}
\hspace*{0.0cm}\textsf{\textbf{(c)}}\\
\vspace*{0.5cm}
\hspace*{0.0cm}{\epsfxsize= 10cm \epsfbox{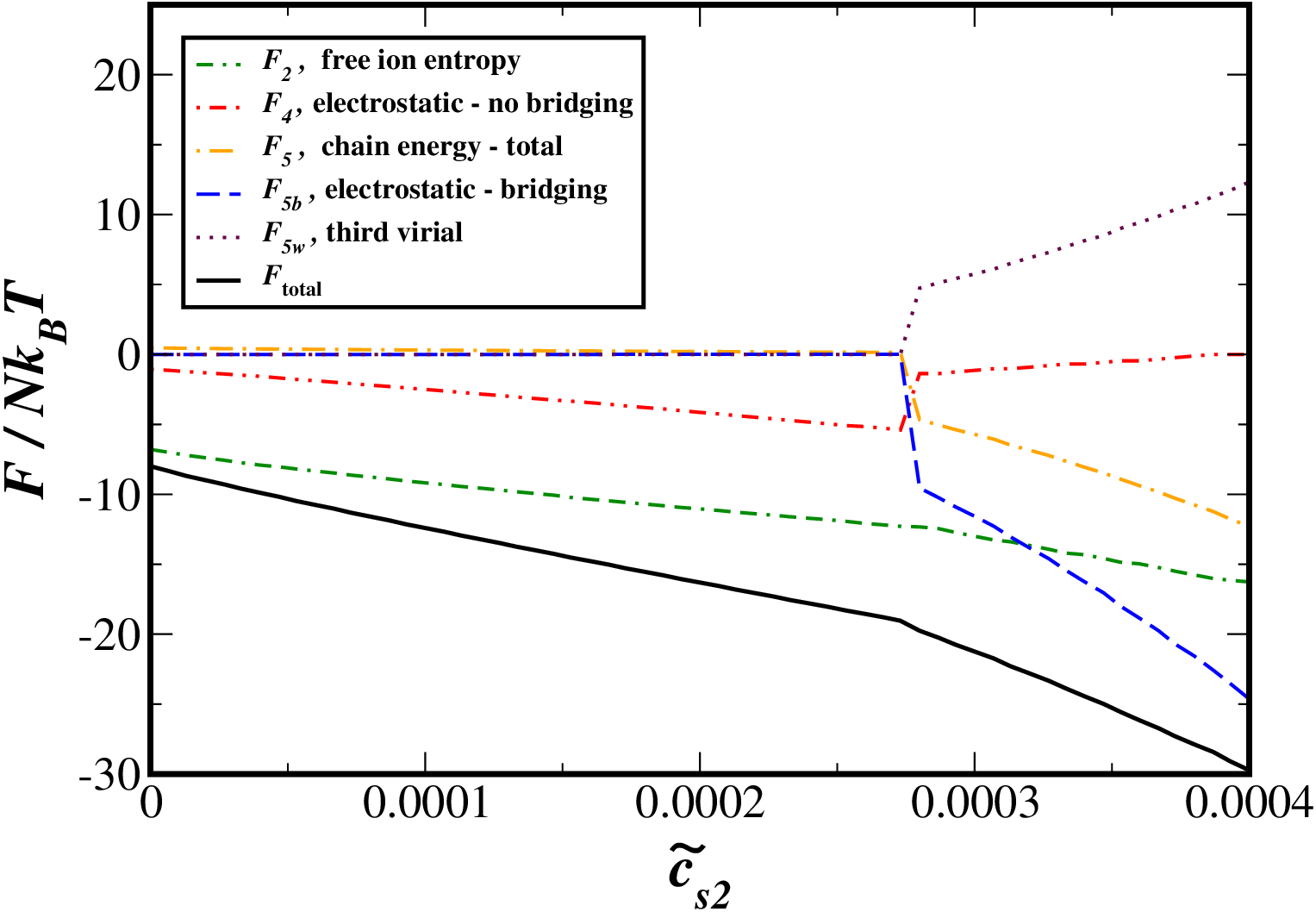}}\\
\caption{Kundagrami \it{et al.}, JCP}
\label{lb-delta-bridge-cs2}
\end{minipage}
\end{figure}

\newpage
\begin{figure}[ht]  \centering
\begin{minipage}{16cm}
\vspace*{-2.0cm}
\hspace*{0.0cm}\textsf{\textbf{(a)}}\\
\vspace*{0.5cm}
\hspace*{0.0cm}{\epsfxsize= 10cm \epsfbox{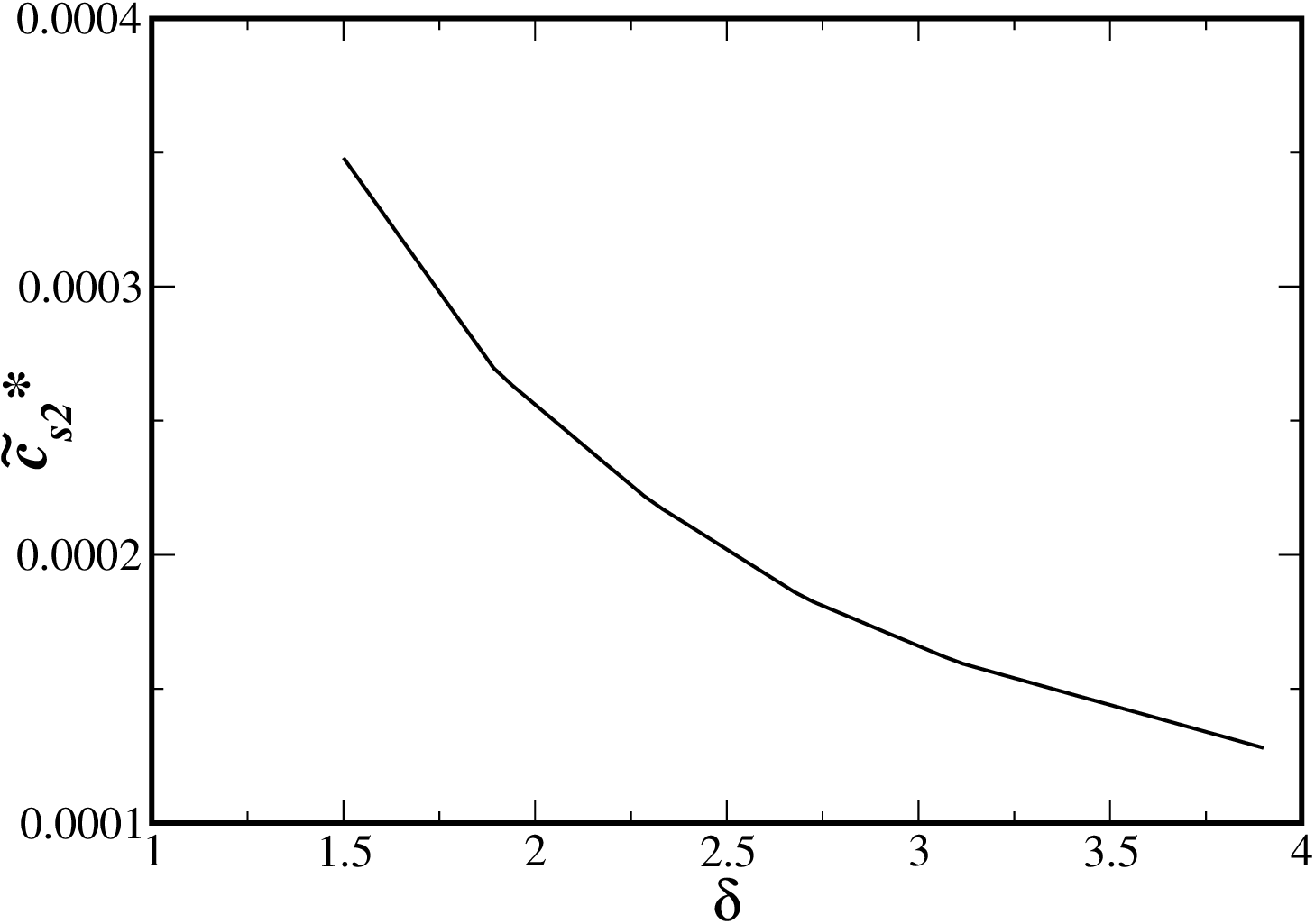}}\\
\vspace*{0.5cm}
\hspace*{0.0cm}\textsf{\textbf{(b)}}\\
\vspace*{0.5cm}
\hspace*{0.0cm}{\epsfxsize= 10cm \epsfbox{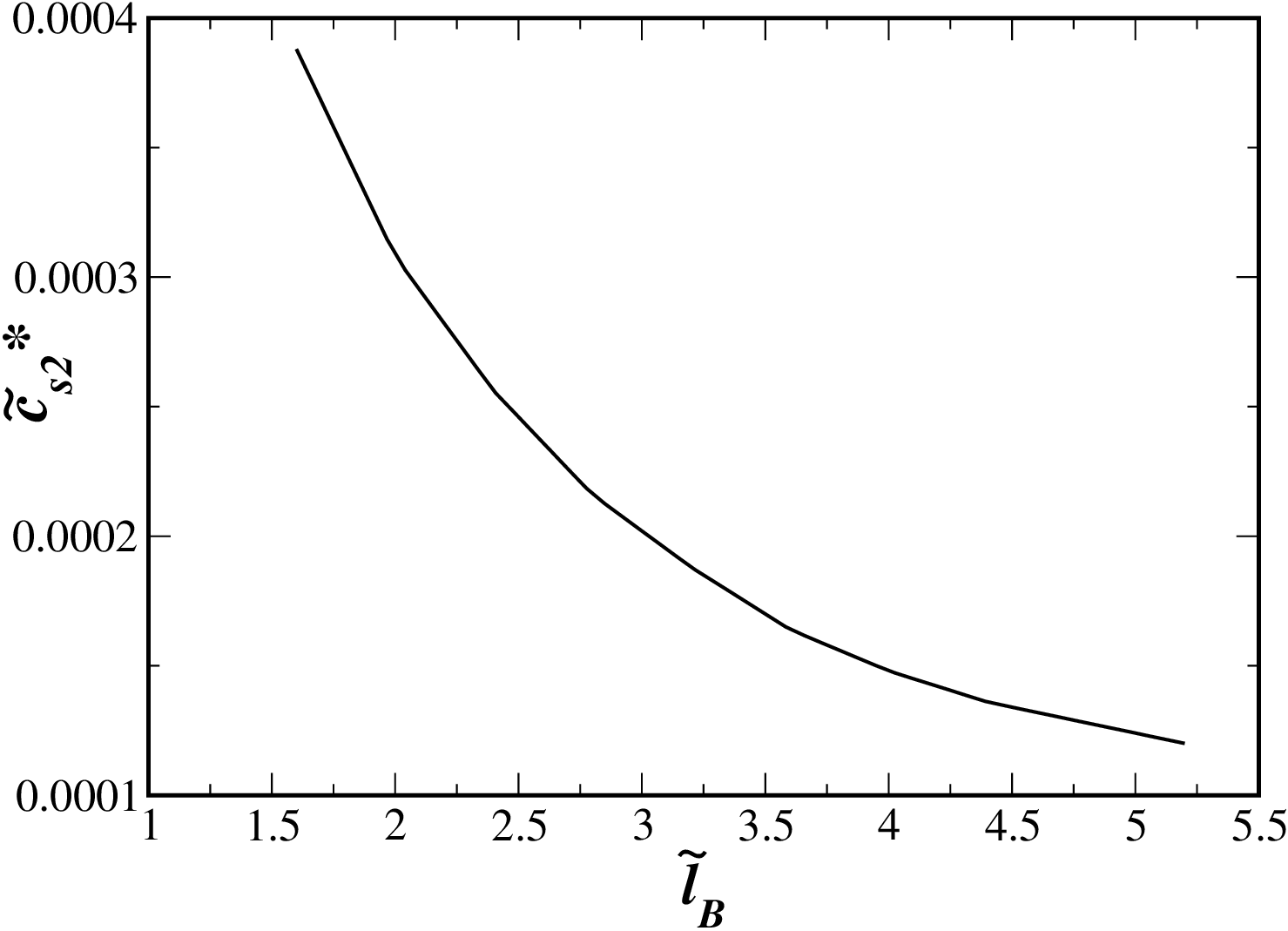}}\\
\vspace*{0.5cm}
\hspace*{0.0cm}\textsf{\textbf{(c)}}\\
\vspace*{0.5cm}
\hspace*{0.0cm}{\epsfxsize= 10cm \epsfbox{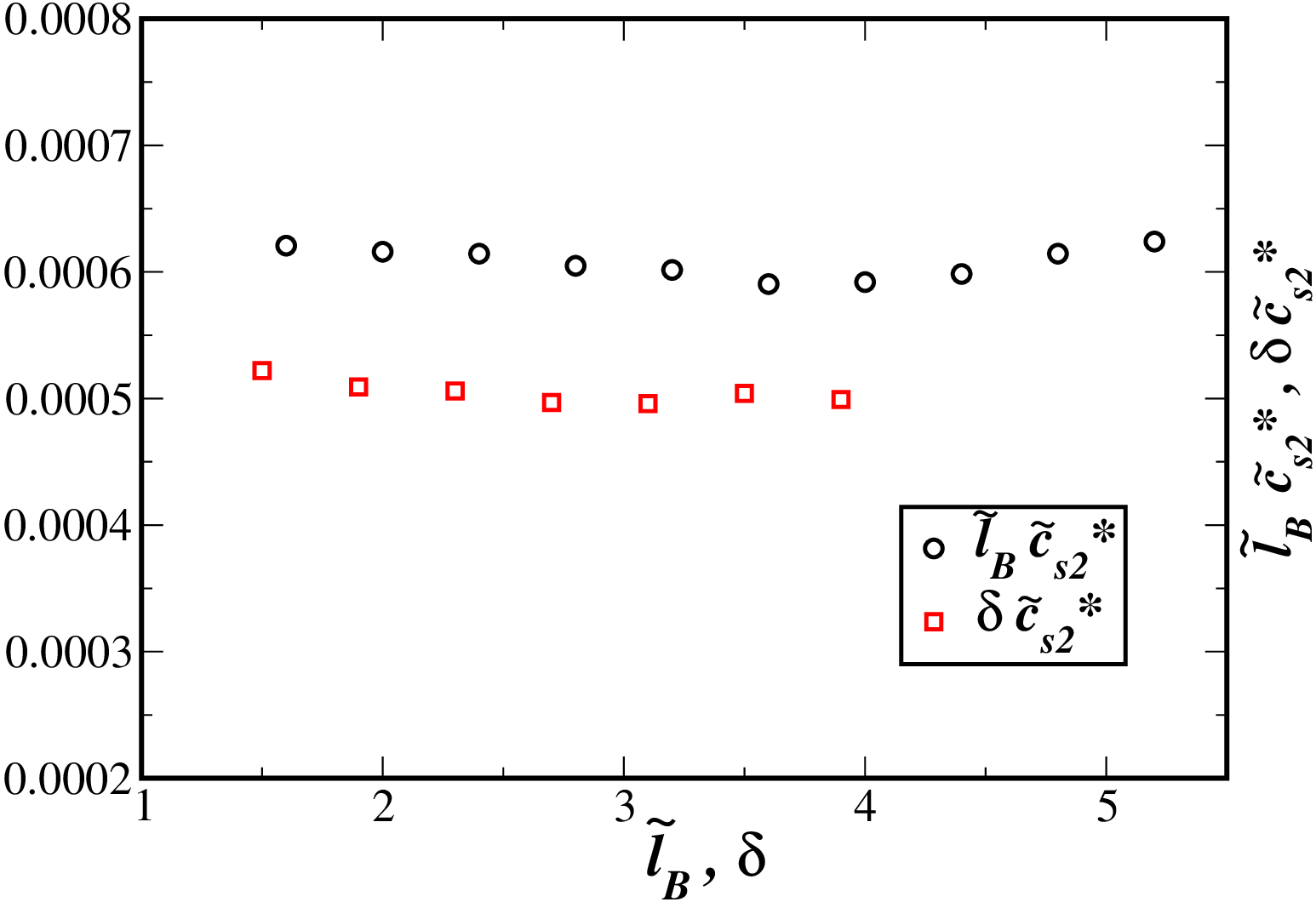}}\\
\caption{Kundagrami \it{et al.}, JCP}
\label{lb-delta-cs2trans}
\end{minipage}
\end{figure}

\end{document}